\documentclass{hch}
\usepackage{graphicx,psfig}
\begin{document}
\title{Supercooled liquids, the glass transition, and computer simulations}
\author{Walter Kob}
\address{Laboratoire des Verres, Universit\'e Montpellier 2, 34095 Montpellier}
\runningtitle{Supercooled liquids, glass transition, and computer simulations}
\maketitle

\section{Introduction}
\label{sec1}

Although in everyday life glasses are mostly associated with pleasurable
or useful things like a glass of wine or window pans, they have been,
and still are, also the topic of research of an impressive number
of investigations. In the last twenty years a particularly strong
effort has been made to solve one of the long standing puzzles
of condensed matter physics: The problem of the glass transition,
i.e. to answer the simple question ``What is a glass?''. The results
of all these experimental, theoretical, and computational efforts is
that today we have a much deeper understanding of the properties of
glass forming systems. Nevertheless one must admit that despite the
impressive activities in the field, some of the key questions are
still not answered and therefore this subject is still a very active
domain of research~\cite{heraclion,alicante,vigo,hernonissos}. The
goal of these lecture notes is therefore to serve as a simple
introduction to the field, to familiarize the reader with some
theories of the glass transition, and to discuss the results of
some computer simulations that have been done to obtain a better
understanding of glass-forming systems. Of course it will not be
possible to present an exhaustive coverage of these different fields,
since they are by now way to vast. But fortunately this is not really
necessary because there exist quite a number of excellent textbooks and
review articles that discuss the various aspects of glassy systems in more
detail~\cite{douglas72,angell81,zallen83,binder86,jackle86,uhlmann86,mezard87,gotze89,fisher91,gotze92,zarzycki91,feltz93,mohanty95,poole95,debenedetti97,young98,cates00,debenedetti01,donth02}.
Thus the goal of the present text is rather to provide a relatively
concise introduction to these various fields and to allow the reader to
familiarize him/her-self with this rapidly evolving topic.

In the first section we hence give an introduction to the dynamics of
supercooled liquids and the glass transition. The second section is
devoted to discuss some of the theoretical approaches used to describe
these systems. Since computer simulations are one of the important current
methods to study glass-forming materials, the following section will be
devoted to discuss the advantages and disadvantages of simulations of
such systems. Finally we will review some results of computer simulations
of glass-forming liquids and discuss to what extend they can be used
to check the validity of theoretical approaches and to increase our
understanding of the static and dynamic properties of glassy materials.

\section{Supercooled liquids and the glass transition: Important facts
and concepts}
\label{sec2}

The goal of this section is to give an introduction to some of the
relevant properties of glass forming systems and to explain some of the
pertinent concepts that are useful to characterize them. This overview
is of course by no means exhaustive but it should nevertheless allow the
reader to get familiar with these kind of systems. For a more exhaustive
description we refer the reader to the additional literature mentioned
in the text.

\begin{figure}[hbtp]
\centering
\includegraphics[width=10.5truecm]{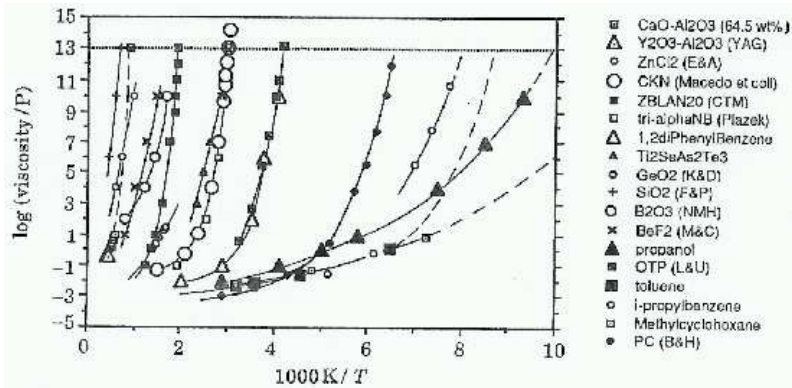}
\caption[]{Temperature dependence of the viscosity of various
glass-forming materials.  Reproduced from Ref.~\cite{angell94} with
permission.}
\label{fig_eta_vs_T}
\end{figure}

Let us consider a system in its liquid state. At sufficiently high
temperatures it can be expected that the viscosity $\eta$ is small, the
diffusion constant $D$ of the atoms (or more general of the constituent
particles) is high, and that the typical relaxation time $\tau$ is
microscopic, i.e. is on the order of a typical vibrational period of the
system which for an atomic liquid is on the order of 0.1-1ps (see below
for a precise definition of $\tau$). If the system is cooled below its
melting temperature $T_m$ one can anticipate it to undergo a static phase
transition, i.e. that it crystallizes. However, in practice it is found
that most liquids can be supercooled to some extend, i.e. it is possible
to study their properties in the (metastable) supercooled regime. (More
details on the lifetime of this metastable state are given below.) Many
experiments as well as computer simulations have shown that the {\it
structural} as well as the {\it thermodynamic} properties of supercooled
liquids show only a relatively weak temperature dependence and that
this dependence can often be extrapolated smoothly from the data above
$T_m$. This is not the case for most {\it dynamic} properties, such as,
e.g., quantities like the viscosity or the diffusion constant. Instead it
is found that these properties usually show a $T-$dependence that is much
more pronounced than the one that would be expected from the one for the
liquid above $T_m$.  As a typical example for such a strong $T-$dependence
we show in Fig.~\ref{fig_eta_vs_T} an Arrhenius plot of the viscosity
$\eta$. From this plot we recognize that a relatively modest change in
temperature (depending on the material between 20\% to a factor of 3)
leads to an increase of $\eta$ by about 12-14 decades. Note that the data
shown covers a wide range of material, including oxides such as SiO$_2$,
as well as molecular liquids such as toluene. A similar behavior is also
found for most polymeric systems. This shows that this dramatic slowing
down of the dynamics is a very general phenomenon.

\begin{figure}[hbtp]
\centering
\includegraphics[width=9.truecm]{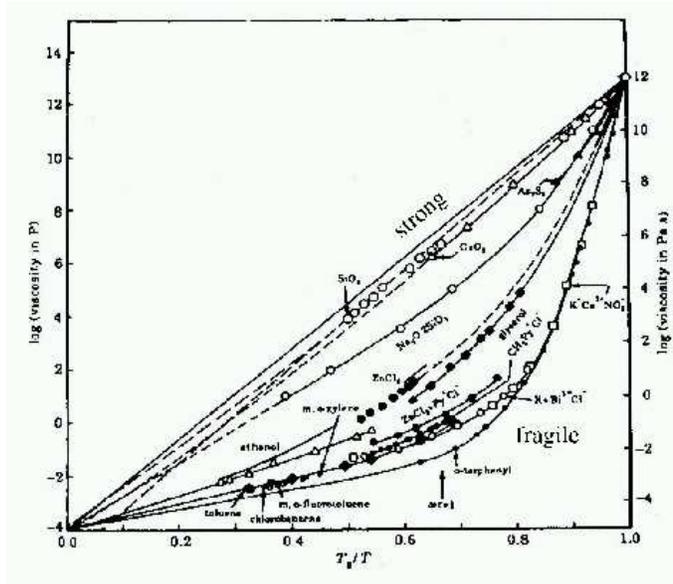}
\caption[]{Viscosity of various glass-forming liquids as a function
of $T_{\rm g}/T$, where $T_{\rm g}$ is the glass transition temperature defined
via $\eta(T_{\rm g})= 10^{13}$~Poise.  Reproduced from Ref.~\cite{angell94}
with permission.}
\label{fig_angell}
\end{figure}

The materials shown in Fig.~\ref{fig_eta_vs_T} have of course
characteristic temperatures (melting point, etc.) that are very different
and hence the different curves spread over a wide range of temperature. It
is therefore useful to make a plot in which one tries to use a reduced
temperature scale. One possibility to do this, proposed first by Laughlin
and Uhlman, is to define a temperature $T_{\rm g}$ at which the viscosity of the
system has the (somewhat arbitrary) value $10^{13}$ Poise (=$10^{12}$Pa s)
and to plot the viscosity as a function of $T_{\rm g}/T$. An example of such a
presentation of the data is shown in Fig.~\ref{fig_angell}~\cite{angell85}
and is commonly called ``Angell-plot''. We see that in this type
of plot the curves for the different materials seem to show a
relatively simple pattern: There are liquids for which $\eta(T)$ is
to a very good approximation just an Arrhenius law (top curves in the
diagram). A prototype of such a material is SiO$_2$ who shows in the
whole accessible temperature range this $T-$dependence. If one moves
downwards in the diagram, one finds materials whose viscosity shows a
bending at intermediate values of $T_{\rm g}/T$. Finally the bottom curves
show a quite pronounced curvature at a temperature around $T_{\rm g}/T\approx
0.7$. Note that each curve can be parametrized in the form $\eta(T) =
\eta_0 \exp(E(t)/k_BT)$, by definition of $E(t)$, and hence the local
slope of the curves can be interpreted as a (temperature dependent)
activation energy $E(t)$. Hence one concludes from the figure that there
are system for which this activation energy is basically independent of
temperature and others for which it increases rapidly with decreasing
$T$. This is evidence that for the first type of systems the mechanism
related to the relaxation of the liquid is independent of temperature,
whereas for the latter type it depends on $T$. Hence Angell coined
the terms ``strong'' and ``fragile'' to distinguish these two types of
behaviors~\cite{angell85}. One possibility to characterize ``fragility''
in a {\it quantitative} way is to consider the slope of $\log (\eta(T))$
vs. $T_{\rm g}/T$ at $T_{\rm g}$: Large slopes correspond to fragile
glass-formers and small ones to strong glass-formers. Although presently
it is not very clear what distinguishes strong and fragile glass-formers
on a microscopic level, it has been found empirically that there is
a significant correlation between fragility and other properties of
the material ($T-$dependence of the specific heat, time dependence of
relaxation dynamics, etc.)~\cite{boehmer93}. Furthermore one observes
the trend that the structure of strong glass-formers is often given by
a relatively open network (e.g. in the case of silica by corner-shared
tetrahedra) whereas the structure of fragile systems is more often
compact, such as ub a hard sphere system. Hence one can conclude that
the fragility is a quantity that does have some physical significance
and later on we will come come back to this point.

Starting from Fig.~\ref{fig_angell} it is now possible to formulate some of
the pertinent questions in the field of glass-forming materials. The
first one is clearly that one wants to understand what the reason is
for the dramatic slowing down of the dynamics. As mentioned above,
all the structural quantities investigated so far do not show any sign
of a unusual $T-$dependence. This is in contrast, e.g., to the case of
second order phase transitions where the slowing down of the dynamics
upon approach to the critical point is closely related to the presence
of a divergent length scale~\cite{stanley71}. Thus for the present time it
seems necessarily to look for another mechanism and in Sec.~\ref{sec3}
we will discuss different theoretical approaches. Note that it is not
even clear whether there is only {\it one} mechanism or whether there are
several ones. E.g. it might well be that the slowing down at small and
intermediate $\eta$ is governed by one mechanism and that at high $\eta$
a different mechanism becomes important. Such a crossover scenario might
e.g. be used to rationalize the bending seen in the viscosity data for
fragile systems.

A further important question, which is related to the first one,
is the exact $T-$dependence of $\eta(T)$ and whether or not
this dependence is the same for other typical time scales of the
system, such as the diffusion constant or the relaxation time. It
is found that at sufficiently high temperatures most liquids show
an Arrhenius dependence. At intermediate and low temperatures the
data can often, but not always!, be fitted well by the so-called
Vogel-Fulcher(-Tammann)-law~\cite{vogel21,fulcher25,tammann26} which
has the form

\begin{equation}
\eta(T)=\eta_0 \exp(A/(T-T_0)) \quad .
\label{eq1}
\end{equation}

\noindent
Thus this functional form predicts a $T-$dependence that for
temperatures close to $T_0$, a temperature that is usually called
``Vogel-temperature'', is significantly stronger than a simple
Arrhenius law. (Note that the latter corresponds to the special case
$T_0=0$.) Although this type of fit gives a good representation of the
data, there is no theoretical foundation for this Ansatz. Nevertheless
it is very useful since it allows for a simple parametrization of the
data with a quite good accuracy. (We point out, however, that there
is experimental evidence that the Vogel-Fulcher-law does not hold
exactly~\cite{stickel95}.)

For intermediate values of $\eta$, i.e. $10^{-1} \mbox{P} \leq \eta \leq 10^2
\mbox{P}$, one often finds that the data is also very well compatible
with a power-law of the form:

\begin{equation}
\eta(T) = \eta_0 (T-T_c)^{-\gamma} \quad .
\label{eq2}
\end{equation}

\noindent
This $T-$dependence is one of the major predictions of the so-called
mode-coupling theory of the glass transition (MCT)~\cite{gotze89,gotze92}
and in Sec.~\ref{sec3} we will discuss this theory in more detail. The
value of the ``critical temperature'' $T_c$ is often found to be around
20-30\% above $T_{\rm g}$. Since Eq.~(\ref{eq2}) predicts a divergence of
the viscosity at $T_c \geq T_{\rm g}$ it is clear that this functional
form cannot be correct for temperatures close to $T_{\rm g}$. Below
we will show however, that for temperatures that are around
$T_c$, the theory is able to rationalize many dynamical properties of
supercooled liquids.

If the Vogel-Fulcher-law given by Eq.~(\ref{eq1}) would indeed hold
down to the Vogel temperature $T_0$ one would have a true divergence
of the viscosity at a {\it finite} temperature. Whether or not such a
divergence really exists is one further fundamental questions in the
field of glass-forming liquids. If the answer is positive it means that
below $T_0$ the system is in an {\it ideal} glass state, i.e. it is
in the global minimum of the free energy, {\it if one considers only
amorphous states and neglects the crystal}. Note that unfortunately
no experiment will be able to give an answer to this question since in
practice it is not possible to equilibrate a real system at a temperature
close to $T_0$ as can be seen as follows: Every experiment last only a
finite time $t_{\rm exp}$ and hence it will not be possible to study the
equilibrium properties of the system below a temperature $T_{\rm g}^{\rm
exp}$ where $T_{\rm g}^{\rm exp}$ is given implicitly by the relation
$\tau(T_{\rm g}^{\rm exp})= t_{\rm exp}$. Since $\tau(T)$ will show at
$T_0$ the same divergence as $\eta(T)$ we have $T_{\rm g}^{\rm exp} >
T_0$. Thus the above posed question can be answered only analytically or
(at least in principle) by computer simulations (see Sec.~\ref{sec4}).

\begin{figure}[hbtp]
\centering
\includegraphics[width=9.truecm]{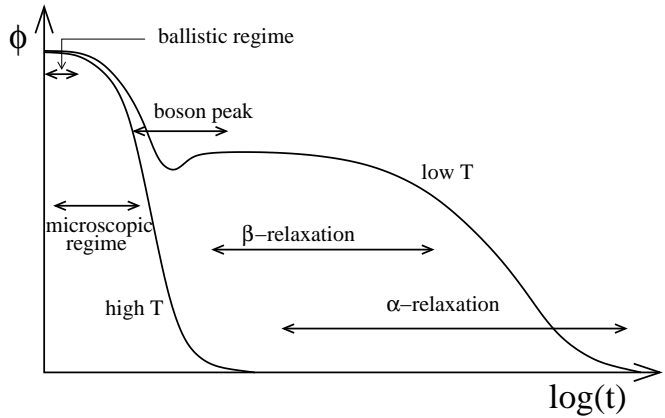}
\caption[]{
Time dependence of a typical correlation function $\Phi(t)$. The two
curves correspond to a temperature that is relatively high, and a
temperature in which the relaxation dynamics of the system is already
very glassy.}
\label{fig_correlator_schematic}
\end{figure}

So far we have discussed the temperature dependence of macroscopic
quantities, like the viscosity. However, many experiments, such as dynamic
light scattering, inelastic neutron scattering, dielectric measurements,
etc.) give also direct access to the time dependence of microscopic
correlation functions, such as the density-density correlator. Since
very often the mentioned macroscopic quantities can be expressed as
the time integral over such correlation functions, the latter certainly
contain more information as the former. (Examples are the viscosity that
is related to the integral over the stress-stress correlation function
or the diffusion constant that is related to the integral over the
velocity correlation function of a tagged particle~\cite{hansen86,balucani94}).
In Fig.~\ref{fig_correlator_schematic} we show in a schematic way
the time dependence of a typical time correlation function $\Phi(t)$
(e.g. the intermediate scattering function $F(q,t)$ discussed in more
detail in sections~\ref{sec3} and \ref{sec5}). The two curves correspond
to two temperatures: One at which the system is in its normal liquid
state and one at which it relaxes only slowly. At the high temperature
the relaxation is relatively simple: At very short times, i.e. at times
much shorter than the typical microscopic times, the time dependence is
quadratic in $t$. This follows directly from the Taylor expansion of the
equation of motion for the particles~\cite{hansen86,balucani94}. \footnote{Here we
assume that this is a liquid that can be described by Newton's equations
of motion. For dissipative systems, such as colloidal particles, some of
the statements have to be slightly modified.} Due to this $t^2$ dependence
this time window is often called ``ballistic regime''. For somewhat larger
times the $t-$dependence $\Phi(t)$ is governed by the interactions between
the particles and hence it is called the ``microscopic regime''. (In
the context of the mean squared displacement of a tagged particle, see
Fig.~\ref{fig_msd_lj_nd}, we will discuss this regime in more detail.) For
even longer times, the $t-$dependence of $\Phi(t)$ is approximated well
by an exponential function, i.e. the system shows a Debye-relaxation.

At low temperatures $\Phi(t)$ shows a more complex time dependence. At
short times one finds again the ballistic regime that is followed by the
microscopic regime. In contrast to the correlator at high temperature,
$\Phi(t)$ now shows at intermediate times a plateau. The time
window in which the correlator is close to this plateau is called
the ``$\beta-$relaxation''. Only for times that are much longer (note
the logarithmic time scale!) the correlation function decays to
zero. The time window in which $\Phi(t)$ decays below the plateau
is usually called the ``$\alpha-$relaxation''. Note that the early
part of the $\alpha-$relaxation coincides with the late part of the
$\beta-$relaxation. The physical meaning of the plateau is given by the
so-called ``cage effect''. At low temperatures each particle is surrounded
by neighboring particles that form a temporary cage around it. At very
short times the particles move ballistically and thus the correlation
function shows a $t^2-$dependence. At somewhat longer times the particles
start to interact with their neighbors and the correlation function
enters the microscopic regime. For intermediate times the particles
are trapped by their neighbors and hence the correlation function is
almost constant. Only for much larger times the particles are able to
leave their cage and hence the correlator starts to decay to zero. In
the context of the mean-squared displacement of a tagged particle we
will return to this trapped motion and will discuss it in more detail
(see Fig.~\ref{fig_msd_lj_nd}). In contrast to the case at high $T$, the final
decay of the correlation function is not an exponential. Although the
precise form is not known, it can usually be approximated well by the
so-called Kohlrausch-Williams-Watts function (KWW)~\cite{kohlrausch47,williams80}, often
also called ``stretched exponential'', which has the form

\begin{equation}
\Phi(t) = A \exp\left( -(t/\tau)^\beta\right) \quad .
\label{eq3}
\end{equation}

\noindent
Here $A$ is an amplitude, $\tau$ can be used to define a relaxation
time, and $\beta \leq 1$ is the KWW-exponent. Note that {\it a priori}
all three parameters will depend on the observable as well as on
the temperature considered. However, sometimes it is found that in a
substantial temperature regime the three parameters are independent
of $T$. In such a case a plot of the correlator vs. $t/\tau$ will give
a $T-$independent master curve and therefore one says that the system
obeys the ``time-temperature-superposition principle''. 

The reason why a time correlation function shows at low $T$ a non-Debye
behavior, i.e. that $\beta < 1$, is still a matter of debate. There
are two extreme scenarios~\cite{richert94}: The first one is that due
to the disorder each particle of the system has a slightly different
neighborhood. Hence also the relaxation dynamics (in particular its
relaxation time) will differ from particle to particle.  Therefore in
this scenario, called ``heterogeneous'', the stretching is due to the
sum of many different Debye-laws with different relaxation times. In the
second scenario, called ``homogeneous'', the relaxation dynamics of the
different particles is not that much influenced by the {\it different}
surrounding disorder. Instead the (general) presence of the disorder gives
rise to a non-Debye relaxation for each particle. Understanding which
one of these two extreme cases are seen in a real supercooled liquid is
presently still a matter of research~\cite{sillescu99,ediger00,richert02}.
But what is already quite clear is that the truth is somewhere in between.

\begin{figure}[hbtp]
\centering
\includegraphics[width=9.truecm]{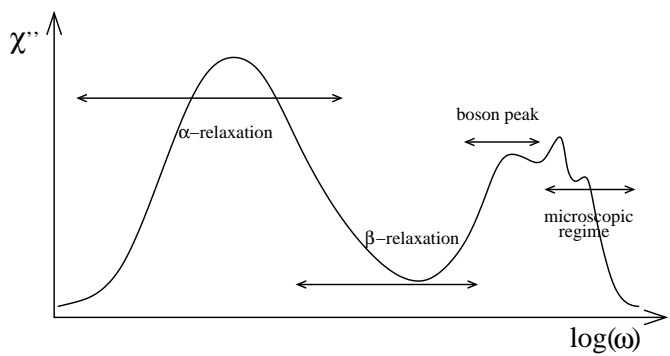}
\caption[]{
Schematic plot of the
frequency dependence of a typical susceptibility $\chi''(\omega)$.}
\label{fig_susceptibility_schematic}
\end{figure}

The last time regime we mention is related to the so-called boson
peak~\cite{winterling75,buchenau86}. In many, but not all, structural
glasses it is found that there exist vibrational excitations that have a
frequency that is about one decade smaller than the typical vibrations in
the system. The precise nature of these excitations is still a matter of
debate~\cite{benassi96,foret96,tara97a,dellanna98,schirmacher98,wischnewski98,rat99,sokolov99,goetze00,hehlen00,masciovecchio00,taraskin00a,grigera01,horbach01,foret02}.
Although they are usually studied in the frequency domain,
where they give rise to a peak, the so-called boson peak (see
Fig.~\ref{fig_susceptibility_schematic}) they also give rise to a feature
in the correlation functions in the time domain, in that the latter show a
small dip after the microscopic regime. For typical glass forming systems,
like e.g. silica, the location of the peak in the frequency domain is
around 1Thz, which thus corresponds to a time scale of 1ps.

The definition of the various time regimes we just gave have
of course also their counterpart in the frequency domain,
i.e. the realm of many experimental techniques (neutron-and
light scattering, dielectric measurements). If one calculates the
time-Fourier transform of a correlation function and multiplies
its imaginary part with $\omega/2k_BT$, one obtains a frequency
dependent susceptibility $\chi''(\omega)$~\cite{hansen86,balucani94}. In
Fig.~\ref{fig_susceptibility_schematic} we show schematically how
$\chi''(\omega)$ depends on the frequency. We recognize that in
general the microscopic regime gives rise to one (or several) peaks in
$\chi''$. The boson peak regime as well as the $\alpha-$relaxation show
up as distinct peaks. (Note however, that depending on how pronounced the
associated excitations are, the former is sometimes only a weak shoulder
on the low-frequency side of the microscopic peaks.) Since in the time
domain the $\alpha-$relaxation quickly moves to larger times if $T$ is decreased,
one finds that in the frequency regime the $\alpha-$peak moves rapidly
to lower frequencies. In contrast to this the microscopic peaks as well
as the boson peak show only a very weak dependence on temperature.

\begin{figure}[hbtp]
\centering
\includegraphics[width=9.truecm]{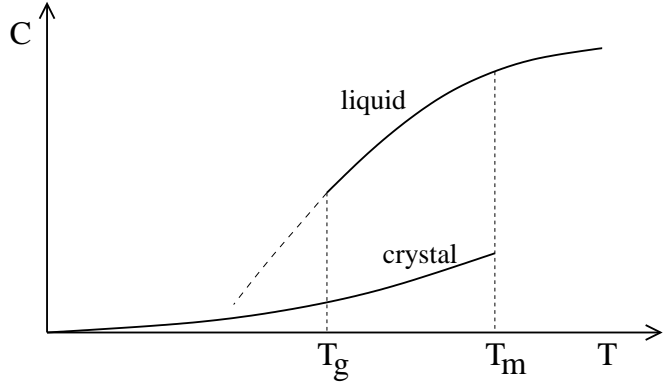}
\caption[]{
Schematic plot of the temperature dependence of the specific heat of 
a glass-forming material in its
liquid-like state and in its crystalline state.}
\label{fig_specific_heat_schematic}
\end{figure}

Having discussed the temperature dependence of structural and dynamical
observables, where the former is weak and the latter is very strong,
we now turn our attention to thermodynamic quantities. Experimentally
it is found that most thermodynamic quantities, such as the pressure,
specific heat, enthalpy, etc. show a very smooth and relatively mild
$T-$dependence. (Note that we talking about the liquid like phase,
i.e. we are not considering non-equilibrium effects that are seen at
the glass transition. These will be discussed below in the context of
Fig.~\ref{fig_volume_schematic_cooling}) In view of this it is surprising
that in 1948 Kauzmann reported that the entropy of supercooled systems
indicate the existence of a phase transition at sufficiently low
temperature and in the following we will review the arguments that
led to this conjecture. In Fig.~\ref{fig_specific_heat_schematic}
we show schematically the temperature dependence of the specific heat
of a glass-forming system. Also included is the specific heat for the
corresponding crystalline system. The latter curve exists only up to
the melting temperature $T_m$ and depends only weakly on $T$, since
anharmonic effects are usually small. The specific heat of the liquid
like branch is higher than the one of the crystal since the liquids is
able to flow, i.e. there are configurational degrees of freedom (those
that give rise to the relaxation/flow of the system) than are not present
in the crystal. Recall that at sufficiently low temperature the typical
time correlation functions of a glass-forming liquid show a separation
of time scales: At short times the particles vibrate in their cage and
only at much longer times they are able to leave this cage and hence
allow the system to flow (see Fig.~\ref{fig_correlator_schematic}). Hence
also the specific heat can be split up into two contribution: One that
is related to the vibrational degrees of freedom and a second part that
is related to the relaxation dynamics (also called ``configurational
degrees of freedom''). (For a more formal approach to this splitting, see
Refs.~\cite{nielsen99,scheidler01}.) The former is quite similar to the
one in a crystal and shows also a similar temperature dependence. Hence
the difference in $C$ between a supercooled liquid and the corresponding
crystal is indeed given by the configurational degrees of freedom.

Using the $T-$dependence of the specific heat it is possible to calculate
the entropy $S(T)$ in the liquid and crystalline phase by means of a thermal
integration, i.e. by using the equality

\begin{equation}
S_\alpha(T_m) = S_\alpha(T) + \int_T^{T_m}dT C/T \qquad \alpha \in \mbox{\{liquid,
crystal\}} \quad .
\label{eq4}
\end{equation}

\begin{figure}[hbtp]
\centering
\includegraphics[width=8.truecm]{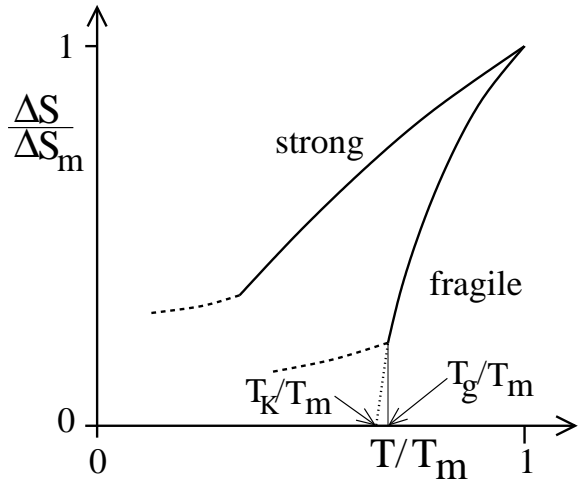}
\caption[]{
Schematic plot of the temperature dependence of the normalized difference
between the entropy of the liquid state and the one of the crystalline
state. The upper and lower curve correspond to the case of strong and
fragile glass-forming material. Note that in the latter case a reasonable
extrapolation of $\Delta S$ seems to vanish at a finite temperature,
the so-called Kauzmann temperature.}
\label{fig_kauzmann}
\end{figure}

\noindent
With this relation it is thus possible to calculate $\Delta S$, the
difference between the entropy of a liquid and the one of the crystal. The
$T-$dependence of this difference, normalized by its value at the melting
temperature, looks schematically as drawn in Fig.~\ref{fig_kauzmann}. We
see that for fragile systems this difference decreases rapidly with
decreasing temperature and a reasonable extrapolation seems to indicate
that it vanishes at a {\it finite} temperature $T_K$, the so-called
Kauzmann temperature. Thus for $T<T_K$ the extrapolation predicts that
the entropy of the glass is below the one of the crystal. Since in
the liquid as well as in the crystal the total entropy is the sum of
the entropy associated with the vibrational degrees of freedom and the
entropy related to the configurational degrees of freedom, $\Delta S=0$
implies that there configurational entropies of the glass and of the
crystal are equal. This is of course a rather surprising result since
the configurational entropy of the (ideal) crystal is zero, since
there exists only one configuration and hence $\Delta S=0$ implies
that there exists also only one configuration for the glass. Hence for
temperatures below $T_K$ the liquid cannot flow anymore, since there
is not state to go to. Thus this final state can be considered as the
``ideal glass''. However, Stillinger has put forward a simple argument that
shows that {\it in a system with short range interactions} the existence
of {\it one} glass state implies immediately that there are exponentially many
(in the system size) and that therefore the configurational entropy of
the disordered state remains always finite~\cite{stillinger88}. Thus one
must conclude that sufficiently close to $T_K$ the extrapolation shown
in Fig.~\ref{fig_kauzmann} is no longer reliable and hence $\Delta S$
does not really vanish. (See also Ref.~\cite{wolfgardt96}.)

It is remarkable that the apparent vanishing of $\Delta S$ at $T_K$ seems
to be related also to the dynamics of the system in that it is found that
often the value of the Vogel temperature $T_0$ from Eq.(\ref{eq1}) is
very close to $T_K$~\cite{richert98}. One possibility for such a connection
is given by the relation proposed by Adams and Gibbs~\cite{adam65},
who used some phenomenological arguments to argue that

\begin{equation}
\tau(T) = A \exp\left(\frac{B}{k_BTS_{\rm conf}}\right) \quad ,
\label{eq5}
\end{equation}

\noindent
where $A$ and $B$ are constants and $S_{\rm conf}$ is the configurational
entropy.  If one makes the Ansatz $S_{\rm conf} \propto T-T_K$,
Eq.~(\ref{eq5}) gives immediately the Vogel-Fulcher-law from
Eq.~(\ref{eq1}) with $T_0=T_K$. 

\begin{figure}[hbtp]
\centering
\includegraphics[width=9.truecm]{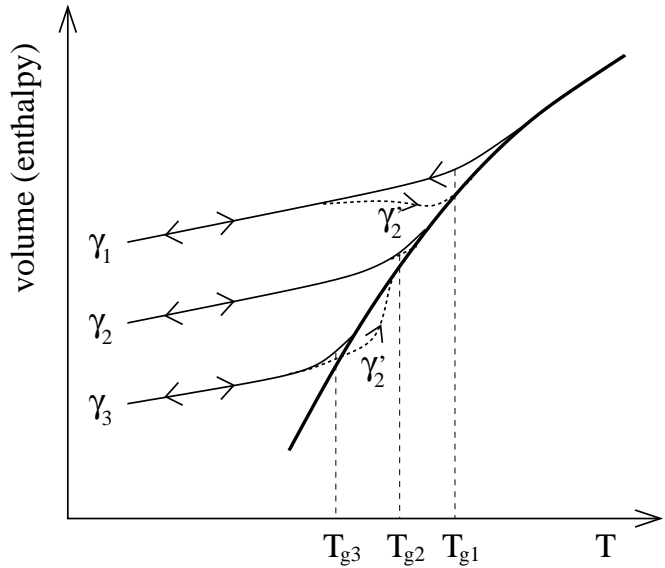}
\caption[]{
Schematic plot of the temperature dependence of the volume (or enthalpy) of a
glass-forming liquid that is cooled with a cooling rate $\gamma_i$ (thin lines) and
reheated with the heating rate $\gamma_2'$ (dotted lines). The bold solid line is the
curve for equilibrium.}
\label{fig_volume_schematic_cooling}
\end{figure}

The results discussed so far concern the behavior of a glass-forming
system in its {\it equilibrium} state (leaving out of course the
possibility of crystallization). However, since the typical relaxation
time of the system increases rapidly with decreasing $T$ it is not
possible to equilibrate the system at arbitrarily low temperatures
since the glass transition intervenes. This is explained schematically
in Fig.~\ref{fig_volume_schematic_cooling} where we show how the
volume (or the enthalpy, or most other thermodynamic quantities)
depend on temperature in a system that is subject to a quench. The
bold solid curve shows the $T-$dependence of the volume {\it in
equilibrium}. Let's assume that we have equilibrated the system at a
high temperature and want to cool it to a lower temperature.  For this
we have to couple it to a heat bath and cool the system with a cooling
rate $\gamma_1$. Since the relaxation time increases with decreasing $T$
there will exist a temperature $T_{\rm g1}$ below which the system is no
longer able to equilibrate on the timescale $1/\gamma_1$ and hence it
will fall out of equilibrium, i.e. it will undergo a glass transition
(thin solid lines in Fig.~\ref{fig_volume_schematic_cooling}). If
we repeat the experiment with a cooling rate $\gamma_2$ the system
will fall out of equilibrium at a temperature $T_{\rm g2}< T_{\rm
g1}$ (see Fig.~\ref{fig_volume_schematic_cooling}). Hence we see
that the glass transition temperature is not a system intrinsic
temperature (such as the melting temperature), since it depends on
experimental parameters like the cooling rate etc. We emphasize
that this dependency of $T_{\rm g}$ is not just of theoretical
interest but is indeed observed in real experiments or computer
simulations~\cite{ritland,yang87,limbach88,johari89,bruning92,bruning94,vollmayr95,vollmayr96,levelut02}.
In Sec.~\ref{sec4} we will come back to this point. In addition it
must be expected that $T_{\rm g}$ depends even on the quantities one
considers, since different observables can have relaxation times that
differ by orders of magnitude thus leading to different glass transition
temperatures.

Since the glass transition temperature depends on the cooling
rate, also the resulting glass will depend on $\gamma$ (see
discussion in context of Figs.~\ref{fig_cooling_rate_density}
and \ref{fig_cooling_rate_dos}). Since the glass is an out of
equilibrium state, all its properties depend on time, i.e. it is
aging~\cite{struik78,mckenna89,cugliandolo02}. E.g. this can be seen
if we start to heat a glass from low to high $T$. This is shown in
Fig.~\ref{fig_volume_schematic_cooling} (dotted lines). Let's start with
the glass that has been produced with a cooling rate $\gamma_1$. If we
heat it with a heating rate $\gamma_2'=\gamma_2< \gamma_1$ it will at
low $T$ follow the cooling curve. However, for temperatures slightly
below $T_{\rm g1}$ the system will already start to equilibrate, since
the available time scale for a change in temperature is higher than the
one the system had during the cooling ($\gamma_2 < \gamma_1$!). Hence the
heating curve will bend downwards toward the equilibrium line. Thus we
conclude that slightly below $T_{\rm g1}$ the properties of the system
will depend on time.

If the initial glass is heated with the same rate as it has been cooled,
the effect that we just mentioned is much less pronounced (see dotted
curve for system $\gamma_2$). Hence the heating curve tracks very closely
the cooling curve and only very closely to $T_{\rm g2}$ one finds a small
aging effect.

If the heating is faster than the cooling, see curve $\gamma_3$ in the
figure where we assume that $\gamma_3< \gamma_2$, the system does not
show pronounced aging until $T_{\rm g3}$. Due to the ``high'' heating
rate it will, however, not be able to start to equilibrate even if it is
at a temperature slightly {\it above} $T_{\rm g3}$, since the relaxation
times are still too large. Hence it will follow a smooth extrapolation
of the ``glass-like'' branch of the volume.  Only when the system has
reached a temperature close to $T_{\rm g2}$ the relaxation times will
be short enough so that the system can equilibrate.

From this discussion it is clear that the falling out of equilibrium is
an effect that depends crucially on the kinetics of the system and hence
on the various time scales of the system and the experiments (cooling
rate, the temperature dependence of the relaxation time and hence on
the observable considered). For technical applications (manufacturing of
glasses, improving the properties of glassy materials) it is important to
have a good understanding of all these kinetic processes and therefore
approximate schemes have been proposed already long time ago to
describe them~\cite{tool31,tool46,narayanaswamy71,moyniham76}. Although
these approaches are useful to characterize the kinetics of the glass
transition, they are of purely phenomenological nature and hence should
not be confounded with the much more recent and advanced theories of
aging~\cite{cugliandolo93,physica_aging,cugliandolo02} despite the fact
that often the same nomenclature is used (such as ``fictive temperature
of the glass''). For more details on these phenomenological theories we
refer the reader to Refs.~\cite{bouchaud98,cugliandolo02}.

\begin{figure}[hbtp]
\centering
\includegraphics[width=9.truecm]{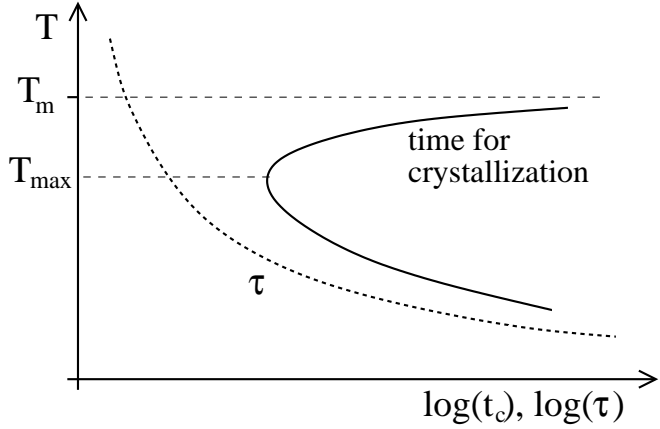}
\caption[]{ 
Schematic plot of the temperature dependence of the time needed for a sample to
crystallize (solid curve). The dashed curve shows the typical behavior of the
relaxation time as a function of supercooling.}
\label{fig_ttt_schematic}
\end{figure}

Everything that we have discussed so far has been made under the
hypothesis that the liquid does not crystallize even in its supercooled
state. Although such an assumption is perfectly valid from a theoretical
point of view, most real supercooled liquids do crystallize after
some time (note that, however, there are liquids that do for all
practical purposes indeed not crystallize, such as atactic polymers). In
Fig.~\ref{fig_ttt_schematic} we plot schematically $t_c$, the time needed
for a liquid to crystallize as a function of its temperature. We see
that if the supercooling is weak, the system can stay in this metastable
state for a long time. Upon further supercooling the driving force
for the nucleation and subsequent crystallization increases and hence
the crystallization time decreases rapidly (note the logarithmic time
axis!). However, in order to form a critical nucleus it is necessary that
the particles move and as we have discussed above this dynamics slows down
quickly with decreasing $T$ (see dashed curve in the figure). Therefore
at low $T$ the nucleation rate starts to decrease and hence the time for
crystallization increases again. Hence we see that there is a temperature
$T_{\rm max}$ at which the crystallization rate has a maximum. Depending
how large this maximal rate is, it may or may not be possible to study
the system experimentally around $T_{\rm max}$. Note, however, that
depending on the details of the shape of the crystallization curve it
might be possible to obtain an equilibrated liquid also {\it below}
$T_{\rm max}$, e.g. by quenching the system sufficiently quickly and
thus passing this critical temperature. We emphasize that this schematic
diagram can also be obtained for real glass-forming systems (see, e.g.,
Ref.~\cite{uhlmann72}) and thus it is not just of theoretical value. One
important consequence of the plot is that it shows that in principle any
liquid/material can be used to form a glass. The only prerequisite is
that one needs an experimental setup to do a sufficiently fast quench.
Finally we remark that since the curves relate temperature, time and
transformation to the crystalline phase, this plot is usually called
``$TTT-$diagram''.

The slowing down of the relaxation dynamics that we have reviewed so
far was due to a change in temperature. It is important to realize,
however, that this is by no means the only possibility to slow down
the dynamics of a system. Other possibilities include an increase of
pressure, evaporation of a solvent in a polymeric system, increasing
the density of crosslinks in polymers, etc. Although these routes
to slow down the system have so far been investigated less than the
temperature route, they are perfectly valid protocols to study the
relaxation dynamics of glass forming systems. As we have mentioned
above, the glass transition is just a kinetic phenomenon in which the
time scale of the experiment becomes comparable with the relaxation
time of the system. It is, however, possible to produce glasses also
without ever going to high temperatures. E.g. one can irradiate a
crystal with neutrons and the so induced radiation damage will slowly
transform the sample into a glass~\cite{klaumunzer92}. The same is
true if one compresses a crystal beyond its stability limits or if
one grinds a crystalline material. Another possibility to form a glass
is to deposit vapor onto a old surface or to make a chemical reaction
(sol-gel transition)~\cite{brinker90,woignier01}. It has to be emphasized
that glasses produced in these different ways will in general differ in
their properties even if some of their macroscopic properties (density,
etc.) are the same. Strictly speaking it is therefore necessary not
only to list some of the properties of a given glass to characterize
the material, but also to specify its complete history of production.

Most of the systems we discussed in this section were liquids in their
supercooled state. We emphasize, however, that a slow relaxation dynamics
is not necessarily related to the {\it supercooled} state. E.g. for the
case of SiO$_2$ the glass transition temperature is $T_{\rm g}=1450$K~\cite{bruckner70}
and the melting temperature is 2000K. From Fig.~\ref{fig_angell} one
recognizes that at $T_{\rm g}/T_m=0.725$ the viscosity has already a value of
$10^7$Poise, i.e. the system is already very viscous (recall that water
at room temperature has a viscosity around $10^{-2}$Poise!). Therefore we
conclude that in order to understand the reason for the slow relaxation
of a system, its supercooled state cannot be a relevant factor. For this
reason one sometimes also uses the term ``glassy liquids'' if one wants to
describe a liquid whose relaxation time is already much larger than the
typical microscopic time scale. In the following we will, however, use
the terms ``supercooled liquids'' and ``glassy liquids'' interchangeably.

Before we end this section we briefly summarize some of of the salient
features of glass-forming systems:

\begin{itemize}
\item
presence of disorder and/or frustration (in the structure, in the 
interactions, \ldots)
\item
no obvious presence of long range order
\item
strong dependence of the relaxation time on an external parameter (temperature, 
pressure, \ldots)
\item
transition to a non-erotic phase
\item
non-exponential decay of the relaxation function at low temperatures
\end{itemize}

Of course this list is neither complete nor is it necessary that
every glass-forming system has all these properties. Instead this list
should give an idea what one typically finds in glass-forming systems
and therefore help to answer the question asked in the Introduction:
``What is a glass''. Using this ``definition'' one realizes that the
class of glass-forming systems is very large: Apart from the atomic
and molecular liquids that we have considered so far, there are the
huge class of polymeric systems that are of immense importance in
our daily life. Of equal relevance are granular materials (sand,
flour, ...) who share many properties of structural glasses at low
temperatures~\cite{cates00,metha94}. Foams and spin glasses are other
systems that share many properties with glass-forming liquids. Last not
least many optimization problems (folding of proteins, traveling salesman
problem, ``\mbox{k-sat} problem'') have a landscape of the cost function
that is very similar to the one of the potential or free energy of more
standard glassy systems and therefore some of the ideas and concepts
used in the later systems are useful in the former as well.

\section{The Mode-Coupling Theory of the Glass Transition}
\label{sec3}

The subject of this section is the above mentioned mode-coupling
theory of the glass transition. Unfortunately the derivation
of all the needed equations is well beyond the scope of
this article and therefore we will refer the reader to the
more specialized literature (see also the review articles in
Refs.~\cite{gotze89,gotze92,schilling94,cummins94,yip95,gotze99}).
Nevertheless, the predictions that MCT makes are quite simple to
understand and therefore we will discuss them in the following. In
addition the theory has the remarkable advantage that it makes detailed
qualitative or even quantitative predictions that can be tested in
experiments or in computer simulations, which is in contrast to most
other theoretical approaches in this field.

The stating point of MCT is the hypothesis that all the structural
properties of a glass-forming liquid are very similar to the one of
the ``normal'', i.e. high temperature, liquid. As we have seen in the
previous section this assumption is well founded. The second important
step in the theory is to exploit the fact that supercooled liquids show a
separation of time scale, i.e. that there are dynamical processes, such
as vibrations, that occur on a microscopic time scale and others, such
as the relaxation, that are observed on a time scale that can be many
orders of magnitude longer than the former one. MCT makes use of this
property of glassy liquids by $i)$ making the choice that the relevant
slowly varying observables are given by the density distribution of
the particles and $ii)$ by deriving equations of motion for these slow
variables. To obtain these equations of motion, the theory makes use
of an {\it exact} formalism that is called the Zwanzig-Mori projection
operator formalism~\cite{mori65,zwanzig61}. Since this formalism is very
useful to understand the structure of the final MCT equations we will
now discuss this approach in somewhat more detail.

\subsection{The Mori-Zwanzig formalism}
\label{sec3.1}

We start with a general classical $N-$particle system. The equation of
motion for an arbitrary phase space function $g$ is given by

\begin{equation}
\dot{g} = i {\cal L} g = \{H,g\} \quad ,
\label{eq6}
\end{equation}

\noindent
where ${\cal L}$ and $H$ are the Liouville operator and the Hamiltonian,
respectively, and $\{.,.\}$ are the Poisson brackets. It is clear that
the set of all possible phase space functions form a vector space and
that one can define a scalar product $(g|h)$ on this space via

\begin{equation}
(g,h) = \langle \delta g^* \delta h \rangle \quad ,
\label{eq7}
\end{equation}

\noindent
where $\langle .\rangle$ is the usual canonical average, and $\delta g=
g-\langle g \rangle$, i.e. is the fluctuating part of $g$. 

Let us assume that thanks to some physical insight we have understood
that the phases space variables $A_n$, $n=1, ...k$ vary only slowly
with time. In the following we will derive equations of motion for
these variables that we will denote by a $k-$dimensional column vector
$\bf{A}$. In a first step we define an operator $\cal P$ that projects
an arbitrary function $g$ on the space spanned by the subset $A_n$:

\begin{equation}
{\cal P}(g) = ({\bf A},g) ({\bf A},{\bf A})^{-1} {\bf A} = 
\sum_{n,m} |A_n) [(A_i,A_j)]_{nm}^{-1} (A_m|f)
\label{eq8}
\end{equation}

\noindent
where $[(A_i,A_j)]_{nm}^{-1}$ is the $n,m$ element of the inverse of the
matrix $(A_i,A_j)$. Furthermore we define the projection operator ${\cal
Q} = 1- {\cal P}$. It is easily verified that $\cal P$ and $\cal Q$
are indeed projectors, i.e. that e.g. ${\cal P}^2={\cal P}$.

From Eq.~(\ref{eq6}) it follows that the time dependence of ${\bf A}$ is
given by ${\bf A}(t)= \exp(i {\cal L}t) {\bf A}$. Inserting the identity
operator ${\cal P} +(1-{\cal P})$ after the propagator $\exp(i{\cal L}t)$
and differentiating this euqation with respect to $t$ one finds

\begin{eqnarray}
d{\bf A}/dt & = & \exp(i{\cal L}t) [{\cal P} + (1-{\cal P})]i{\cal L} {\bf A}\\
            & = & i{\bf \Omega} \cdot {\bf A}(t)+\exp(i {\cal L}t)(1-{\cal P})i{\cal L}{\bf A}
\quad,
\label{eq9}
\end{eqnarray}

\noindent
where we have introduced the so-called frequency matrix $i{\bf \Omega}=
({\bf A}, i {\cal L}{\bf A}) \cdot ({\bf A},{\bf A})^{-1}$. The last
term in Eq.~(\ref{eq9}) can be written as~\cite{balucani94}

\begin{equation}
\exp(i {\cal L}t)(1-{\cal P})i{\cal L}{\bf A} = \int_0^t d\tau \exp[i
{\cal L} (t-\tau)]i {\cal P} {\cal L} {\bf f}(\tau) + {\bf f}(t)
\label{eq10}
\end{equation}

\noindent
where the function ${\bf f}(t)$ is called ``fluctuating force'' (see
below why) and is given by

\begin{equation}
{\bf f}(t) = \exp[i(1-{\cal P}){\cal L}t]i (1-{\cal P}) {\cal L} {\bf A} \quad .
\label{eq11}
\end{equation} 

Note that Eq.~(\ref{eq11}) shows that the time evolution of the function
${\bf f}$ from its initial value $i (1-{\cal P}) {\cal L} {\bf A}$
is ruled by the propagator $\exp[i(1-{\cal P}){\cal L}t]$ instead of
the usual one: $\exp[i{\cal L}t]$. In addition we also see immediately
that due to the factor $(1-{\cal P})$ we have 

\begin{equation}
({\bf A},{\bf f}(t)) = 0 \quad ,
\label{eq11b}
\end{equation}

\noindent
i.e. ${\bf f}(t)$ is always orthogonal to ${\bf A}$.

Using in Eq.~(\ref{eq10}) the relations ${\cal P} {\cal L} {\bf f}(\tau) =
({\bf A}, {\cal L}{\bf f}(\tau)) \cdot ({\bf A}, {\bf A})^{-1} {\bf A}$
and $i({\bf A}, {\cal L}{\bf f}(\tau)) = i ((1-{\cal P}) {\cal L} {\bf
A}, {\bf f}(\tau)) = -({\bf f}(0),{\bf f}(\tau))$, we obtain for the
equation of motion~(\ref{eq9}) the expression

\begin{equation}
\dot{\bf A} = i{\bf \Omega} \cdot {\bf A}(t)-\int_0^t d\tau {\bf M}(\tau) \cdot {\bf
A}(t-\tau) + {\bf f}(t) \quad ,
\label{eq12}
\end{equation}

\noindent
where we have introduced the {\it memory function} ${\bf M}(t)= ({\bf
f}, {\bf f}(t)) \cdot ({\bf A},{\bf A})^{-1}$. The equation of motion
for ${\bf C}(t) = \langle {\bf A}^*(0) {\bf A}(t) \rangle$, the matrix
of the correlation functions, is now easily obtained by multiplying Eq.~(\ref{eq12}) with
${\bf A}$:

\begin{equation}
\dot{\bf C}(t) = i {\bf \Omega} \cdot {\bf C}(t) -\int_0^t d\tau {\bf
M}(\tau) \cdot {\bf C}(t-\tau) \quad.
\label{eq13}
\end{equation}

\noindent
Note that the third term on the right hand side of
Eq.~(\ref{eq12}) does not show up in Eq.~(\ref{eq13}) due to
Eq.~(\ref{eq11b}). Equations~(\ref{eq12}) and (\ref{eq13}) are usually
called {\it generalized Langevin equation} and {\it memory equation},
respectively. Both of them are {\it exact} since no approximation has
been made so far. The formal solution of Eq.~(\ref{eq13}) is obtained
by introducing the Laplace transformed quantities

\begin{equation}
{\bf \widehat{C}}(z) = \int_0^\infty dt \exp(-zt) {\bf C}(t) \quad \mbox{and} \quad
{\bf \widehat{M}}(z) = \int_0^\infty dt \exp(-zt) {\bf M}(t)
\label{eq14}
\end{equation}

\noindent
which give

\begin{equation}
{\bf \widehat{C}}(z) = \left[ z{\bf I} -i {\bf \Omega} + {\bf \widehat{M}}(z) \right]^{-1}
\cdot {\bf \widehat{C}}(0) \quad .
\label{eq15}
\end{equation}

\noindent
Here ${\bf I}$ is the unit matrix. Although this solution is formally
exact, it is not that useful since the time dependence of the memory
function ${\bf M}(t)$ is very complicated.  However, {\it if} the
set ${\bf A}$ of variables that we have chosen includes {\it all}
slowly varying quantities, the memory function must either be a fast
variable, or it must be possible to express it (at least approximately)
as a function of the variables in ${\bf A}$, such as, e.g. as linear
combinations of $A_i^nA_j^m$, with $n,m = 1,2, ...$. This coupling of
the variables (modes) has given rise to the name ``mode-coupling
theory''. We emphasize that the Mori-Zwanzig projection operator
formalism and the mode-coupling approximations are not just a technique
to describe the dynamics of supercooled liquids but a method that
has been widely used in all sort of situations where one is able
to identify pertinent slow variables (critical phenomena, Brownian
motion, etc.)~\cite{forster75,barrat99}. Whether or not this approach
is successful depends of course on the situation of interest and it is
in most cases difficult to tell in advance. However, it has been found
that there are indeed quite a few situations in which this method works
remarkably well.

\subsection{Application of the Mori-Zwanzig formalism to glass-forming
systems}
\label{sec3.2}

Having discussed the general formalism that leads to mode-coupling
equations, we now turn our attention back to the case of supercooled
liquids. For this type of systems it has been proposed that the slow
variables are $\delta \rho({\bf q},t)$, the fluctuations in the density for
wave-vector ${\bf q}$, and hence the mode-coupling equations are equations of
motion for the corresponding correlation functions $F(q,t)$: 

\begin{equation}
\delta \rho({\bf q},t) = \sum_{j=1}^N \exp[i {\bf q}\cdot {\bf r}_j(t)]
\quad \mbox{and} \quad
\quad F(q,t) = \frac{1}{N} \langle \delta \rho({\bf q},t) \delta \rho({\bf q},0)^* \rangle
\label{eq16}
\end{equation}

\noindent
(Note that here we have assumed that the system is isotropic. Hence $F(q,t)$
depends only on the modulus of ${\bf q}$.) Also
important is the fluctuation in the density of a tagged particle $\delta
\rho_s({\bf q},t)$ for wave-vector ${\bf q}$ and its associated correlation function
$F_s(q,t)$:

\begin{equation}
\delta \rho_s({\bf q},t) = \exp[-i {\bf q}\cdot {\bf r}_j(t)]
\quad \mbox{and} \quad
\quad F_s(q,t) = \frac{1}{N} \langle \delta \rho_s({\bf q},t) \delta \rho_s({\bf q},0)^* \rangle
\label{eq16b}
\end{equation}

\noindent
(Note that due to the thermal average in the definition of
$F_s(q,t)$, the correlator does not depend on the particle index
$j$.) The space-time correlation functions $F(q,t)$ and $F_s(q,t)$
are usually called the coherent and incoherent intermediate scattering
functions~\cite{hansen86}. Apart from their theoretical importance,
their significance lies in the fact that they are directly accessible in
inelastic neutron scattering~\cite{hansen86,balucani94,lovesey94}. Hence
obtaining a good understanding of their $q$ and $t-$dependence is
considered a very important task in the theory of liquids (which is not
yet a solved problem).

A somewhat lengthy calculation shows that the MCT equations for
$\Phi(q,t)=F(q,t)/S(q)$, where $S(q)=F(q,0)$ is the static structure
factor, are given by~\cite{bgs84,gotze89}:

\begin{equation}
\ddot{\Phi}(q,t)+ \Omega^2(q)\Phi(q,t)+\int_0^t\left[M^{\rm reg}(q,t-t')+ \Omega^2(q)
M(q,t-t')\right]\dot \Phi(q,t')dt'  = 0\quad .
\label{eq17}
\end{equation}

\noindent
Thus we see that this equation does indeed have the form of the general
mode-coupling equation~(\ref{eq13}). The value of the squared frequency
$\Omega^2(q)$ follows directly from the short time expansion of the
equations of motion for the particles and is given by

\begin{equation}
\Omega^2(q)=q^2 k_B T/(mS(q)) \quad ,
\label{eq18}
\end{equation}

\noindent
where $m$ is the mass of the particles. The function $M^{\rm reg}(q,t)$,
also often called the ``regular part of the memory function, governs
the time dependence of $F(q,t)$ at short (i.e. microscopic) times,
e.g. just after the ballistic regime. Often it is modeled by a Gaussian
Ansatz~\cite{tankeshwar87,tankeshwar95}, but its precise functional form
is presently not really known. Note that this part of the memory function
is present also in ``normal'' liquids (i.e. liquids that are not glassy)
where it governs the relaxation dynamics and thus obtaining an accurate
form of the $M^{\rm reg}(q,t)$ is a current problem of standard liquid
state theory.

The long time behavior of $\Phi(q,t)$ is ruled by the memory function
$M(q,t)$. Within the approximation of MCT this kernel is a quadratic
function in $\Phi(q,t)$:

\begin{equation}
M(q,t)=\frac{1}{2 (2\pi)^3}\int d{\bf k}
V^{(2)}(q,k,|{\bf q}-{\bf k}|)\Phi(k,t) \Phi(|{\bf q}-{\bf k}|,t) 
\label{eq19}
\end{equation}
where the vertex $V^{(2)}$ is given by
\begin{equation}
V^{(2)}(q,k,|{\bf q}-{\bf k}|)=\frac{n}{q^2} S(q)S(k)S(|{\bf q}-{\bf k}|)
\left(\frac{{\bf q}}{q}\left[ {\bf k}c(k)+
({\bf q}-{\bf k}) c(|{\bf q}-{\bf k}|) \right] \right)^2 \quad .
\label{eq20}
\end{equation}

\noindent
Here $n=V/N$ is the particles density and $c(k)$ is the so-called
direct correlation function that is related to the structure
factor via $c(k)= n(1-1/S(q))$~\cite{hansen86}. Hence we see that
Eq.~(\ref{eq17})-(\ref{eq20}) are a closed set of equations that define
the time dependence of $\Phi(q,t)$ and hence of $F(q,t)$. Thus the
solution of this set of equations are correlation functions that can
in principle be directly compared with data from an experiment or a
computer simulation.

Some remarks on these equations and their solutions:
\begin{itemize}
\item
The form of Eq.~(\ref{eq17}) is the one of a damped harmonic oscillator
with a damping that depends on time. With decreasing temperature
the static structure factor becomes more peaked and hence also the
vertices $V^{(2)}$ become more peaked. This means that the damping
$M(q,t)$ increases, at least for certain wave-vectors, and that hence
the relaxation becomes slower. Note that since the memory function
depends itself on the correlators, see Eq.~(\ref{eq19}), a slowing down
of the latter will increase the former, therefore leading to an even
slower dynamics for $\Phi(q,t)$. This non-linear feedback effect leads to
a very strong dependence of the relaxation dynamics on the structural
quantities and hence gives a qualitative explanation why the relaxation
times of glass-forming systems change so quickly as a function of external
parameters (temperature, pressure, etc.).

\item
It can be shown that if the vertices are sufficiently large, the time
scale at which the correlators start to decay to zero increases beyond
any bound. This means that there is a critical temperature $T_c$ (or
pressure $p_c$) at which the correlators do not decay to zero anymore and
hence at which the system undergoes a transition from an ergodic state
to a non-ergodic one. Often this is called an ``ideal glass transition''
or the ``mode-coupling transition''. Due to this ideal transition the
equations are also called ``ideal mode-coupling equations'' and at the end of
this section we will briefly discuss their generalization.

\item
The only input needed to solve these equations are the static quantities
$S(q)$, $m$, $n$ and $T$. In addition there is the memory function $M^{\rm
reg}(q,t)$ for the dynamics at short times. However, it can be shown
that the details of $M^{\rm reg}(q,t)$ do neither affect the existence
of the MCT-transition nor the value $T_c$ at which it occurs. Also the
$t-$dependence of $F(q,t)$ at intermediate and long times, i.e. times much
longer than the microscopic times, is independent of $M^{\rm reg}(q,t)$,
{\it apart from a system universal shift in the time scale}. Hence MCT
predicts that the relaxation dynamics is completely independent of the
microscopic dynamics, apart from a overall change of the time scale.

\item
The equations of motion discussed so far concerned only the
auto-correlation function for the coherent intermediate scattering
function $F(q,t)$. However, it is possible to derive also equations
of motion for the incoherent function $F_s(q,t)$, defined in
Eq.~(\ref{eq16b}). One finds that these equations have the following form:

\begin{equation}
\ddot{F}_s(q,t)+ \frac{q^2 k_BT}{m}F_s(q,t)+
\int_0^t\left[M^{\rm reg,s}(q,t-t')+ 
M^s(q,t-t')\right]\dot F_s(q,t')dt'  = 0\quad .
\label{eq21} 
\end{equation}

Here the memory function is given by

\begin{equation}
M^s(q,t) = \frac{nk_B T}{(2\pi)^3m} \int d {\bf q}' \left(
\frac{{\bf k}'\cdot {\bf q}}{q}\right)^2 c(q') S(q') \Phi(q',t) F_s(|{\bf q}-{\bf
q}'|,t)
\label{eq22}
\end{equation}

Thus the memory function for the dynamics of $F_s(q,t)$ contains the
normlized coherent intermediate scattering function $\Phi(q,t)$ and
hence we see that in order to obtain the time dependence of $F_s(q,t)$,
we need first the one for $F(q,t)$. This result is quite reasonable, since
$F_s(q,t)$ describes the motion of a tagged particle in its environment
(which is disordered and time dependent). Hence it must be expected
that a relevant input for the time dependence of $F_s(q,t)$ is the time
dependence of the relative motion of the tagged particle with respect to
its surrounding, and $F(q,t)$ provides exactly this kind of information.

\item
The vertices $V^{(2)}$ as given by Eq.~(\ref{eq20}) depend only on
the static structure factor $S(q)$. Strictly speaking this is not
quite true, since there is also an additional term that depends on
the three-particle correlation function known as $c_3({\bf q},{\bf
k})$~\cite{hansen86,barrat88,denton89}. However, for {\it simple}
liquids it has been found that this additional contribution to
$V^{(2)}$ influences the quantitative predictions of the theory only
very weakly~\cite{barrat90,sciortino01} and hence it is in most cases
neglected. For the case of liquids that have a structure that is given
by an open network, this is no longer true and hence these additional
terms have to be taken into account~\cite{sciortino01}. In context with
Figs.~\ref{fig_nep_lj} and \ref{fig_nep_sio2} we will come back to this point.

\item
So far we have only discussed one-component atomic liquids. It is
relatively straightforward to generalize the MCT equations also to
multicomponent systems and one finds that in that case the equations of
motion become matrix-equations for the partial intermediate scattering
functions~\cite{barrat90}.

It is also possible to take into account the situation that
two (or more) particles are permanently linked together, which
opens the door to obtain a description for the translational
and orientational correlation functions in molecular and polymeric
systems~\cite{schilling97,kawasaki97,franosch97,chong98,chong98b,fabbian99,gotze00,letz00,chong01,theenhaus01,chong02}.
The equations one obtains are, however, rather complicated and therefore
difficult to solve numerically.  However, if one is interested only in
simple orientational correlation functions, it is possible to simplify
these equations considerably and hence to treat such molecular systems
also numerically~\cite{chong98,chong98b,chong01,chong02}.

\end{itemize}

The equations~(\ref{eq17})-(\ref{eq20}) are very complicated and
hence it is not surprising that no analytical solution is known. But
even obtaining a qualitative understanding of these solutions is a
formidable task and therefore it would be nice to have some means
to simplify these equations. One possibility for doing this has been
proposed in a seminal paper by Bengtzelius {\it et al.}~\cite{bgs84}
(see also Leutheusser~\cite{leutheusser84}). By solving numerically
the MCT-equations for the case of a hard sphere system, these authors
found that the main contribution to the memory function $M(q,t)$
comes from wave-vectors that are close to the peak in the static
structure factor. Hence they proposed to approximate the $S(q)$
by a single $\delta-$function located at this peak: $S(q)= \zeta
\delta(q-q_0)$. Here $\zeta$ is the strength of the $\delta-$function
and thus is assumed to increase with decreasing temperature. It is now
found immediately that the MCT-equations simplify significantly in that they 
boil down to a {\it single} equation of the form:

\begin{equation}
\ddot{\phi}(t)+ \Omega^2\phi(t)+
\zeta \Omega^2 \int_0^t\phi^2(t-t')\dot \phi(t')dt'  = 0\quad ,
\label{eq23}
\end{equation}

\noindent
where $\phi(t)= F(q_0,t)/S(q_0)$ and we have neglected the regular part
of the memory function $M^{\rm reg}$, since, as discussed above, it
does not influence the solution of the equations qualitatively. If one
approximates the structure factor by the sum of two $\delta-$functions,
one obtains two equations of motion in which the memory function is a
simple polynomial of the two correlators. The equations of motion obtained
in this way are called ``schematic models'' and Eq.~(\ref{eq23}) is a
particular example. The interest in such models lies in the fact that
the generic features of their solutions are the same as the ones of the
full MCT-equations. However, since they are significantly more simple that
these full equations, they are very useful to obtain a general overview on
the possible $t-$dependence of the solutions. In addition they have also
been found to be very useful to discuss experimental data such as spectra
of dynamical light scattering~\cite{alba95,franosch97b,ruffle99,voigtmann00}.

Before we start to discuss the generic properties of the solutions of
the MCT-equations we briefly mention a very interesting connection
of these equations with the time-correlation functions of spin
glasses~\cite{binder86,young98}. From the way the MCT-equations
are derived it is clear that for the case of a real liquid they are
only an approximation. Since this approximation is non-perturbative
(there is no small parameter), the resulting equations have often
been criticized as being ``uncontrolled''. Although strictly speaking
this might be true for the case of real liquids, it has been found
that the predictions of the theory are often so remarkably good, see
Refs.~\cite{gotze92,cummins94,gotze99} for reviews and some examples
will be discussed below, that the observed agreement cannot be a mere
accident. Nevertheless it would of course be comfortable if one would know
systems for which the MCT-equations are exact. Surprisingly such systems
do exist, as has been shown in a series of seminal papers by Kirkpatrick,
Thirumalai and Wolynes~\cite{kirkpatrick87,kirkpatrick88,thirumalai88}.
These authors found that in certain type of mean-field spin glasses
(Potts glasses and $p-$spin models) the equation of motion for
the spin-auto-correlation function obeys exactly the MCT-equations
as given by the schematic models. Thus this surprising connection
between the structural glasses and spin glasses gave evidence that
these two type of systems, which at a first glance have very little
in common, might be much more similar than has been anticipated. In
recent years a large effort has been made to test to what extend this
connection holds, i.e., to test whether concepts and ideas that work
well for one type of system (overlap of states~\cite{parisi02},
violation of the fluctuation dissipation theorem in the glassy
phase~\cite{bouchaud98,cugliandolo02}, driving of the system, calculation
of the Kauzmann temperature~\cite{mezard98,coluzzi00,mezard00,coluzzi02}, etc.) are
also useful for the other type of disordered system. All these activities
resulted that the field is currently still extremely active and highly
interesting.

We now turn our attention to the discussion of some of the pertinent
predictions of MCT. One of the most important prediction is the one
already mentioned above, namely that there exists a critical temperature
$T_c$ at which the $\alpha-$relaxation time of the system diverges. All the
other predictions that will be presented in the following hold, strictly
speaking, only very close to $T_c$, i.e. it is assumed that the parameter
$\sigma = (T_c-T)/T_c$ is small. How small in practice $\sigma$ has to be
so that the predictions of the theory hold is presently not very clear
and depends also on which prediction one considers~\cite{franosch97c,fuchs98}.
The results of experiments and computer simulations have shown, however,
that often it is sufficient that $|\sigma|$ is less than 0.1-0.2, i.e 
the predictions of the theory can be seen in a reasonable interval
of $T$.

MCT predicts not only that the $\alpha-$relaxation time diverges at $T_c$,
but also gives the specific form of this divergence, namely a power-law:

\begin{equation}
\tau_x(T) = C_x (T-T_c)^{-\gamma} \quad .
\label{eq24}
\end{equation}

\noindent
Here $\tau_x(T)$ denotes the $\alpha-$relaxation time of a correlator
that we label with $x$ (e.g. this might be the relaxation time for
the coherent intermediate scattering function for wave-vector $q$, the
incoherent function, etc.). $C_x$ is a prefactor that depends on $x$ and
whose $T-$dependence close to $T_c$ can be neglected, and $\gamma$ is a
{\it system universal constant}, i.e. it is independent of $x$. Hence the
implication of Eq.~(\ref{eq24}) is that all correlation functions will
show a divergence of the relaxation time at the {\it same} temperature
$T_c$ and also the critical exponent $\gamma$ will be independent of the
correlator considered. (Note that by ``all correlators'' we mean here
that the overlap of the observable of interest with $\delta \rho({\bf q},t)$
does not vanish.) One consequence of this result is that also the inverse of
the diffusion constant $D$ should show a power-law divergence, since $D$ can be
expressed as the limit $q\to 0$ of the incoherent intermediate scattering
function. Hence we have:

\begin{equation}
D^{-1}(T) \propto \tau_x(T) = C_x (T-T_c)^{-\gamma} \quad .
\label{eq24b}
\end{equation}

If the MCT equations are solved numerically, it is found
that with increasing coupling (i.e. decreasing temperature
in Eqs.~(\ref{eq17})-(\ref{eq20}), or increasing $\zeta$ in
Eq.~(\ref{eq23})) the time-dependence  of the solution changes
from a shape as given qualitatively by the high $T$ curve shown
in Fig.~\ref{fig_correlator_schematic} to a shape as the one of the
low $T$ curve in this figure. Thus we see that the theory is able to
describe the cage effect discussed in the previous section at least
qualitatively. The theory goes, however, much further in that it does
not only give a qualitative description of the cage effect but predicts
that in the $\beta-$relaxation regime the time dependence of all
correlation functions (``all'' in the sense described above) is universal
in the following sense: Let $\Phi_x(t)$ be an arbitrary correlator. Then
it is predicted that close to the plateau $\Phi_x(t)$ can be written as

\begin{equation}
\Phi_x(t)= \Phi_x^c + h_x G(t)  \quad ,
\label{eq25}
\end{equation}

\noindent
where $\Phi_x^c$ is the height of the plateau (often also called
``non-ergodicity parameter'', since it gives the fraction of the memory of
the initial state that still exists in the time window of the plateau),
$h_x$ is an amplitude, and $G(t)$ is a system universal function. Thus
we see that the theory predicts that $\Phi_x(t)-\Phi_x^c$ is just the
product of a $x-$dependent amplitude and a $x-$independent function of
time. Therefore the relation given by Eq.~(\ref{eq25}) is called the
``factorization property''. Furthermore the theory predicts that the
function $G(t)$ has a very specific dependence on temperature in that
it can be written in the form

\begin{equation}
G(t) = \sqrt{|\sigma|} \, g_\pm (t/t_\sigma), \qquad \sigma= (T_c-T)/T_c \quad ,
\label{eq26}
\end{equation}

\noindent
where the $T-$independent functions $g_\pm$ correspond to $T<T_c$ and
$T>T_c$, respectively. The time scale $t_\sigma$ is the location of
the plateau and is predicted to show a power-law divergence of the form
$t_\sigma= t_0|\sigma|^{1/2a}$, where $t_0$ is a microscopic time scale.

The time dependence of the functions $g_\pm$ can be obtained by solving
(numerically) the following non-linear equations~\cite{gotze85}:

\begin{equation}
\mp z^{-1} + z g_\pm^2(z) + i \lambda \int_0^\infty d \tau \exp(i z
\tau)g_\pm ^2(\tau) = 0.
\label{eq27}
\end{equation}

\noindent
Here the constant $\lambda$, which is also often called ``exponent
parameter'', can be calculated from the vertices $V^{(2)}$. Since the
functions $g_\pm(t)$ give the time dependence of the correlation functions
in the $\beta-$relaxation regime, they are called ``$\beta-$correlator''.

The solution of Eq.~(\ref{eq27}) are not known analytically but it can
be shown that they have the following asymptotic form (i.e. very close
to the plateau):

\begin{equation}
g_\pm(\hat{t}) = \hat{t}^{-a} \quad \mbox{for} \quad \hat{t} \ll 1, \quad \mbox{i.e.} 
\quad t_0 \ll t \ll t_\sigma
\label{eq28}
\end{equation}

\noindent
This functional form is called ``critical decay''. For times much larger than
$t_\sigma$ the form of $g_-(t)$ is given by

\begin{equation}
g_-(\hat{t}) = B \hat{t}^b  \quad \mbox{for} \quad \hat{t}\gg 1 \quad \mbox{i.e.} 
\quad t_\sigma \ll t \ll \tau \quad ,
\label{eq29}
\end{equation}

\noindent
a functional form that in this context is called ``von Schweidler
law''~\cite{gotze85}. The exponents $a$ and $b$ from Eqs.~(\ref{eq28})
and (\ref{eq29}) can be calculated from the exponent parameter $\lambda$ from
Eq.(\ref{eq27}) via

\begin{equation}
\frac{\Gamma(1-a)^2}{\Gamma(1-2a)}=\frac{\Gamma(1+b)^2}{\Gamma(1+2b)} = \lambda
\quad ,
\label{eq30}
\end{equation}

\noindent
where $\Gamma(x)$ is the usual $\Gamma-$function. Furthermore the theory make
the interesting prediction that the exponent $\gamma$ of the power-law for the
$\alpha-$relaxation time $\tau$, see Eq.~(\ref{eq24}), is related to $a$ and
$b$ by means of the relation

\begin{equation}
\gamma= \frac{1}{2a}+ \frac{1}{2b} \quad .
\label{eq31}
\end{equation}

\noindent
Hence we see that once the exponent parameter $\lambda$ is known, the
exponents $a$, $b$ and $\gamma$ are determined.

Note that according to MCT the $\beta-$relaxation is not just the
cross-over regime from the microscopic dynamics to the relaxation dynamics
at long times, but it is a process {\it of its own} (that is, however,
tightly connected with the $\alpha-$process). To understand what we mean
by this we consider the imaginary part of a dynamical susceptibility which
was shown schematically in Fig.~\ref{fig_susceptibility_schematic}. In
the context of that figure we have said that with decreasing temperature
the $\alpha-$relaxation peak moves quickly to the left, whereas the
microscopic peak shows only a very weak $T-$dependence. Thus the existence
of a minimum between the two peaks is of course trivial. The non-trivial
statement of MCT is that close to the minimum $\chi''(\omega)$ is not
just the sum of the low-frequency part of the microscopic peak and the
high-frequency part of the $\alpha-$peak, but that there is an additional
intensity, the $\beta-$process that is described by the $\beta-$correlator
occurring in Eq.~(\ref{eq25}).

Having discussed some of the predictions of MCT for the
$\beta-$relaxation, we now turn our attention to the $\alpha-$relaxation,
i.e. the time window in which the correlators fall below the plateau. In
the context of Eq.~(\ref{eq24}) have have already discussed one important
property of this relaxation regime, namely that the relaxation times
diverge at the same critical temperature and that also the exponent is
the same. A further prediction of the theory is that the time-temperature
superposition principles (TTSP), mentioned in Sec.~\ref{sec2}, should
hold. This means that at a temperature $T$ the time dependence of a
correlator $\Phi(t,T)$ can be written as follows:

\begin{equation}
\Phi(t,T) = \hat{\Phi}(t/\tau(T)) \qquad ,
\label{eq32}
\end{equation}

\noindent
where $\hat{\Phi}$ is a master function, and $\tau(T)$ is the usual
$\alpha-$relaxation time. (Note that if the TTSP holds, the precise
definition of $\tau(T)$ is irrelevant for Eq.~(\ref{eq32}) to hold.)

As mentioned in Sec.~\ref{sec2}, the shape of the correlators is usually
approximated well by the KWW function given in Eq.~(\ref{eq3}). Although
this function is not an {\it exact} solution of the MCT equations, it
has been found that also the numerical solutions of the MCT equations
are approximated very well by a KWW function. Hence it can be concluded
that MCT is predicting the stretching of the time correlation function.

Before we end this section on MCT it is appropriate to make some
comments to what extend this theory is applicable to real glass-forming
systems. As mentioned above, the MCT equations are {\it exact} for
certain mean-field models (see articles by Parisi and Cugliandolo in these
proceedings~\cite{parisi02,cugliandolo02}). For short range systems the
equations will of course only be an approximation and the real question is
how good this approximation really is. Unfortunately there is presently
no clear answer. There are short range spin glasses, e.g.  the 10 state
Potts model, whose static and dynamic properties are very different from
the mean-field prediction, whereas in other spin glasses the mean-field
prediction seems to be very reliable~\cite{brangian02,marinari98}. For
the case of structural glasses there is a large body of literature in
which the applicability of MCT has been investigated for all sort of
systems (colloids, gels, molecular glasses, network-forming glasses,
polymers, etc.) and it has been found that in general the theory does
a remarkably good job in explaining many dynamical features of these
systems on a qualitative level, or sometimes even quantitatively (see
Ref.~\cite{gotze99} for a recent review of the experimental literature).

There are, however, also results from experiments or simulations
that seem to be at odds with the predictions of the theory. A careful
inspection of such findings shows that either the theory has not been
used in an appropriate way (e.g. because the authors have not fully
understood the theory) or that only the {\it asymptotic} predictions of
the theory have been tested. By ``asymptotic predictions'' we mean
the following: As mentioned above, most predictions of the theory,
and in particular the ones presented in this article, hold only very
close to the critical point $T_c$. These predictions are nothing else
than the mathematical properties of the solutions of the MCT equations,
i.e. they are independent of whether whether or not these equations
correctly describe a real system. If the parameter $\sigma=(T-T_c)/T_c$
cannot be considered as small anymore, these properties no longer
hold and therefore the properties of the solutions change, and in
particular they cannot be cast into simple forms as given, e.g., by
Eqs~(\ref{eq28}) or (\ref{eq29}). Although even in such a case it is
possible to make some analytical predictions~\cite{franosch97c,fuchs98} that
can be tested~\cite{gleim00}, it is in most cases simpler to make a direct
comparison with the numerical solution of the MCT equations. However,
this approach is quite involved and therefore not often done. But in
principle it is the only systematic procedure if one really wants to
find out whether or not the MCT equations give a good description of
the relaxation dynamics of a real system.

Apart from the just mentioned problem on the applicability of the MCT
equations, there is, however, a much more important one, namely the
fact that the MCT equations, as presented here, predict a divergence
of the relaxation time at a finite temperature. In Sec.~\ref{sec2}
we have seen that there is experimental evidence that there does
indeed exists a {\it finite} temperature at which $\tau$ diverges,
the Vogel temperature $T_0$ (see Eq.~(\ref{eq1})), which is close to
the Kauzmann temperature $T_K$ in which thermodynamic quantities seem
to indicate a transition (see Fig.~\ref{fig_kauzmann}). However, in
experiments it is found that the divergence of $\tau$ close to $T_0$ is
given by the Vogel-Fulcher law, Eq.~(\ref{eq1}), and not by a power-law
as predicted by MCT. Thus this is evidence that one should probably
not identify $T_0$ with $T_c$. Furthermore the exact calculations for
the mean-field models have shown that these models do indeed have two
transitions~\cite{kirkpatrick87,kirkpatrick88,thirumalai88,gross85,kirkpatrick95}.
A {\it thermodynamic} one at low temperature $T_{\rm s}$ and a {\it
dynamic} one at a higher temperature $T_D$. (This latter transition is
due to the fact that at $T_D$ the configuration space splits up into
domains that are separated by infinitely high barriers.) If we recall
that the MCT equations for these spin models describe the ergodic $\to$
non-ergodic transition at $T_D$, we thus are tempted to make the analogy
that the Kauzmann temperature of structural glasses corresponds to the
temperature $T_s$ in spin glasses and that the $T_c$ corresponds to the
temperature $T_D$. Since real systems are not mean-field, the mentioned
barriers that are present the mean-field models are no longer infinite
and hence there is no longer a sharp transition. Instead one can expect
that close to $T_c$ the dynamics changes from a {\it flow-like one}, to
one in which a few particles {\it hop} in a cooperative way over local
barriers and recent investigation on the properties of the landscape
of supercooled liquids have given evidence that this is indeed the
case~\cite{broderix00,angelani00,buchner00,cavagna01,grigera02}. Hence
we come to the conclusion that the sharp transition predicted by MCT at
$T_c$ is avoided and one finds instead only a change of the transport
mechanism for the particles. Depending on the system, this change may
be very pronounced, such as in colloidal suspension where close to
$T_c$ (or in that case close to the critical packing fraction of the
particles) the relaxation times become macroscopically large, or only
relatively mild, such as in the case of strong glass-forming liquids
(silica, B$_2$O$_3$, glycerol) were only a relatively mild change
occurs~\cite{wuttke95,franosch97b,horbach99}.

It is of course a legitimate question to ask whether or not MCT is
indeed not able to describe the relaxation dynamics also below $T_c$. In
principle the Mori-Zwanzig formalism allows to obtain also equations
of motion for that situation and the ``only'' thing one has to do is
to take into account further slow variables~\cite{gotze87,gotze88}. The
result of such a calculation are equations of motion in which the memory
function has a somewhat more complicated form, and these equations
are usually called ``generalized MCT equations'' or ``MCT-equations
including the hopping terms'' since they include the ergodicity
restoring hopping terms (or at least some of these terms) (see also
Ref.~\cite{das86}). However, so far it has not been possible to come
up with an {\it explicit} expression for this memory function, i.e. one
that does not depend strongly on the approximations that one is forced
to make. Therefore only the qualitative form of this generalized memory
function is known and if one wants to use these equations of motion to
describe, e.g., real data for $F(q,t)$, quite a few fit parameters are
left~\cite{cummins93}. Despite this difficulty it is, however, possible to
make at least some predictions that should be valid independently of the
functional form of the generalized memory function. The most important
one is that the sharp transition at $T_c$ no longer exists and that instead
there is a cross-over from a power-law dependence above $T_c$ to an weaker
$T-$dependence (e.g. Arrhenius) below $T_c$. Depending on the strength of
the ``hopping parameters'' in the generalized equations, this transition
can be very sharp, or very smooth which allows to distinguish between
fragile and strong glass formers. Furthermore it can be shown that despite
the presence of the hopping processes, the factorization property of the
$\beta-$relaxation regime still holds, although in a somewhat different
form as the one given by Eq.~(\ref{eq25})~\cite{gotze87,gotze88}. Thus
we see that even if our current ignorance of the exact form of the
generalized MCT-equations does not allow to make a quantitative
comparison between the prediction of the theory and the real data,
certain qualitative checks are still possible. In the past such tests
have indeed been made~\cite{baschnagel95,horbach01b} and it has been found that the
theory does indeed allow to rationalize some of the dynamical properties
of glass forming liquids also at temperatures below $T_c$.

Thus we can conclude that the theory is able to give a good description
of the relaxation dynamics not only above $T_c$ but is also a useful
theory in a certain temperature range below $T_c$. In that sense MCT is a
very valuable theoretical contribution that has helped to understand the
relaxation dynamics of glass-forming liquids at intermediate temperatures
\footnote{Note that in experiments of glass-forming liquids with strong
or intermediate fragility it is often found that $T_c$ is 20\%--30\%
above $T_{\rm g}$. For the strong glass-former silica there is evidence
from simulations and experiments that that $T_c$ is even a factor of two
higher than $T_{\rm g}$~\cite{horbach01b,hess96,rossler98,horbach99}. In
view of these results it is therefore more appropriate to say that the
theory works well at temperatures in which the viscosity has values
between $10^{-1}$Poise and $10^3$Poise.}. In this $T-$range the theory
has identified the existence of a cross-over in the transport mechanism of
the particles with many interesting and universal features. Nevertheless,
to what extend the theory is also able to give a correct description of
the relaxation dynamics of atomic liquids at temperatures at which the
viscosity is much higher, is presently not really known and the focus
of current research.

\section{Computer Simulations of glass-forming Systems}
\label{sec4}

As we have already mentioned in the introduction, computer simulations
are one of the important tools to investigate the properties of
glass-forming systems.  The goal of this section is therefore to
discuss this method in some detail.  However, in the following we
are not going to give a general introduction to computer simulations
since a much more extended coverage can be found in several excellent
textbooks~\cite{allen90,allen93,frenkel96,binder96,binder97,landau00}.  Instead we
will discuss mainly some of the technical issues of such simulations that
are of particular importance for glass-forming systems. But even with this
restriction the field is still to vast to give a comprehensive review and
therefore we will leave out almost completely important topics like the
simulation of spin glasses etc. and focus instead on structural glasses.

Roughly speaking there are two broad classes of simulations: {\it
Molecular dynamics simulations} and {\it Monte Carlo simulations}, and in
the following we will discuss these two methods in somewhat more detail.

In a molecular dynamics (MD) simulation one tries to solve Newton's
equations of motion for the system of interest and thus to obtain
the trajectory of the particles in configuration space.  From these
trajectories many properties of the system can be directly calculated,
at least in principle. The problem with MD is that that the known numerical
methods to solve these equations work only reliably for a time scale that
is significantly smaller ($O(10^{-2})$) than the shortest typical time
scale of the system, which in a liquid is given by the local motion of
the particles and hence is on the order of 0.1ps. One well known example
for such an algorithm is due to Verlet and is given by:

\begin{eqnarray}
{\bf r}_i(t+h) & = & {\bf r}_i(t) + h{\bf v}_i(t) + \frac{h^2}{2} {\bf F}_i(t)\\
{\bf v}_i(t+h) & = & {\bf v}_i(t) + \frac{h}{2} \left[{\bf F}_i(t)+ {\bf F}_i(t+h)\right]
\quad .
\label{eq33}
\end{eqnarray}

\noindent
Here ${\bf r}_i(t)$ and ${\bf v}_i(t)$ are the position and velocity
of particle $i$ at time $t$, ${\bf F}_i$ is the force acting on this
particle, and $h$ is the time step which for typical liquids has to
be on the order of 1fs. In Sec.~\ref{sec2} we have seen that upon
cooling the relaxation time of a glass-forming liquid increases by
12-15 decades in time. Thus if one wants to investigate this slowing
down by means of a simulation it would be necessary to propagate the
system by means of Eq.~(\ref{eq33}) by the same number of decades, which
is currently completely impossible even for systems that are as small
as 100 particles. The best one can presently do is to simulate a system
with $O(1000)$ particles for $O(10^9)$ time steps, but most present day
simulations cover less than $10^7$ time steps since usually one has to do
many independent runs to improve the statistics, to investigate different
temperatures, to average (in spin glasses) over different realizations
of disorder, etc., all of which together usually takes many years of
single processor CPU time.  Hence we recognize that one of the major problems
in the simulation of glass-forming systems is the huge time range that
should be covered. Note, however, that the propagation of the system in
its configuration space is so slow because the dynamics of the particles
is {\it realistic}. Hence this type of simulations allows to calculate
directly dynamical properties for the system, such as the diffusion
constant, the velocity of sound, the vibrational density of states, and
more. This is the main reason why MD is, despite the mentioned drawback,
a very popular method to study glass-forming systems.

In a Monte Carlo simulation this drawback does, at least in principle,
no longer exist. Since in this method the only rule that has to be
obeyed is the one of detailed balance, i.e. the fact that $w(s\to
s')$, the transition probabilities for the propagation of the system
from an arbitrary state $s$ to another arbitrary state $s'$, fulfill
the relation $w(s\to s')/w(s'\to s) = \exp(-\beta[H(s)-H(s')])$, where
$\beta=1/k_BT$ and $H(s)$ is the value of the Hamiltonian at state $s$,
a clever choice of the function $w(s\to s')$ will allow to equilibrate
the system also at low temperatures. Unfortunately this is true only
in principle. So far it has not been possible to find a updating scheme
that allows to equilibrate an arbitrary off lattice system also at low
temperatures. For certain special cases we do, however, have algorithms
that allow to do this, e.g. systems of hard disks or binary mixtures
of particles with very similar radii~\cite{dress95,grigera01b}. For
other models it is possible to come up with algorithms
that allow at least to probe the system (in equilibrium!) at
relatively low temperatures, such as, e.g. the parallel tempering
method~\cite{swendsen86,marinari92,hukushima96,hansmann97,hukushima98,hansmann99,kob00,yamamoto00}.
Despite this progress one is still quite far away from the ultimate goal:
To have a robust algorithm that is able to equilibrate a large class
of glass-forming systems (having a substantial number of particles)
also at very low temperatures.

Since so far it is not possible to simulate reasonably sized systems over
very large time, and since we have seen that with decreasing temperature
the relaxation times of glass-forming systems increase rapidly, it is
presently difficult to use computer simulations to investigate the {\it
equilibrium} properties of such systems at low $T$. What does ``low''
mean? As mentioned above the length of a typical run is presently
around $10^7$ time steps. With a step size $h$ of a few fs, this gives a
total length of 10-50ns which corresponds to about $10^2$Poise. Using
Fig.~\ref{fig_angell} we thus recognize that present day simulations
are just barely able to see the upturn in the $\eta(T)$ curves seen
experimentally in fragile liquids.

\begin{figure}[hbtp]
\centering
\includegraphics[width=9.truecm]{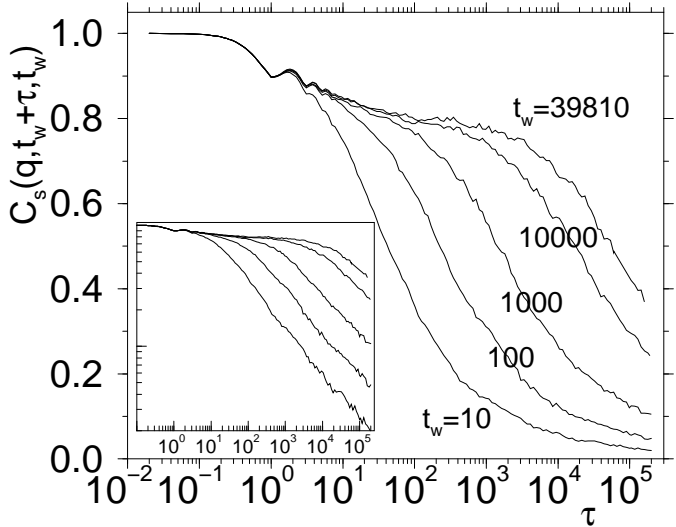}
\caption[]{
Main figure: Time dependence of the correlation function
$C_s(q,t_w+\tau,t_w)$ defined in Eq.~\protect(\ref{eq35}) for a binary
Lennard-Jones mixture that has been quenched from a temperature $T_i=5.0$
to a temperature $T_f=0.4$. The different curves correspond to the stated
waiting times $t_w$. Inset: The same data in a log-log plot showing hat
at long times the correlator shows a power-law dependence on time. Adapted
from Ref.~\cite{kob97}.}
\label{fig_C_q_t_tw}
\end{figure}

Of course one might be tempted to circumvent this problem by making a
relatively rapid quench to low temperatures, ``anneal'' the system for
some time, and then to start to measure the properties of the system
(structure, diffusion constant, relaxation dynamics, etc.). Although
such an approach sounds quite reasonable, a systematic investigation of
such a protocol shows that the so obtained results have very little to
do with the ones of the system in {\it true} equilibrium and therefore
should not be done. To demonstrate this one can, e.g., investigate the
time-dependence of the intermediate scattering function $F_s(q,t)$ (see
Eq.~(\ref{eq16b}) after such a quench. {\it In equilibrium} $F_s(q,t)$
depends only on the time difference $t$, i.e.

\begin{equation}
F_s(q,t)=\langle \rho_s({\bf q},t) \rho_s^*({\bf q},0)\rangle = 
\langle \rho_s({\bf q},t+\tau) \rho_s^*({\bf q},\tau)\rangle \quad.
\label{eq34}
\end{equation}

In the out-of-equilibrium case the second equality does no longer hold 
and therefore it is necessary to keep track of both time arguments $t$
and $\tau$, e.g., we have to generalize $F_s(q,t)$ to the function

\begin{equation}
C_s(q,t_w+\tau,t_w)= \langle \rho_s({\bf q},t_w+\tau) \rho_s({\bf q},t_w)\rangle \quad ,
\label{eq35}
\end{equation}

\noindent
where $t_w$ is the so-called ``waiting time'', i.e. the time span
between the start of the measurement and the time at which the system has
been driven out of equilibrium. In Fig.~\ref{fig_C_q_t_tw} we show the
time dependence of a $C_s(q,t_w+\tau,t_w)$ for a binary Lennard-Jones
system that at time zero has been quenched from a relatively high
temperature to a low temperature~\cite{kob97}. The different curves
correspond to different waiting times $t_w$. We see that, in contrast
of the situation in equilibrium, the relaxation dynamics of the
system depends strongly on the waiting time in that the dynamics
slows down for increasing $t_w$. One thus says that the system is
``aging''. Although the existence of this phenomenon has been known
since quite some time~\cite{struik78}, it is only in recent years that
attempts have been made to describe it by means of statistical mechanics
concepts~\cite{ciuchi88,cugliandolo93,cugliandolo94,bouchaud95,kurchan96,physica_aging,monthus96,bouchaud98,nieuwenhuizen98,latz00,latz01,cugliandolo02}.
Note that the figure shows not only that the typical time scale for the
decay of the correlation function depends on $t_w$, but also that the
shape of the function has changed, since (see inset) it is compatible
with a power-law whereas in equilibrium it can be approximated well by
a KWW-function (see Sec.~\ref{sec2}). Hence we conclude that for the
investigation of the {\it equilibrium} dynamics of a slowly relaxing
system it is most important to check carefully that one has indeed
reached equilibrium. Note that experience has shown that time correlation
functions are much more sensitive to out-of-equilibrium effects than
static quantities. This can be understood by recalling that most static
quantities show only a very weak $T-$dependence, whereas dynamical
observables show a very strong one (see Sec.~\ref{sec2}). Therefore
in most cases it is not advisable to monitor just quantities like the
energy, pressure etc. in order to check whether or not a system has
reached equilibrium. Much more useful are the time correlation functions
like the one defined in Eq.~(\ref{eq35}) or the function that is the
analogous generalization of the mean-squared displacement.

\begin{figure}[hbtp]
\centering
\includegraphics[width=9.truecm]{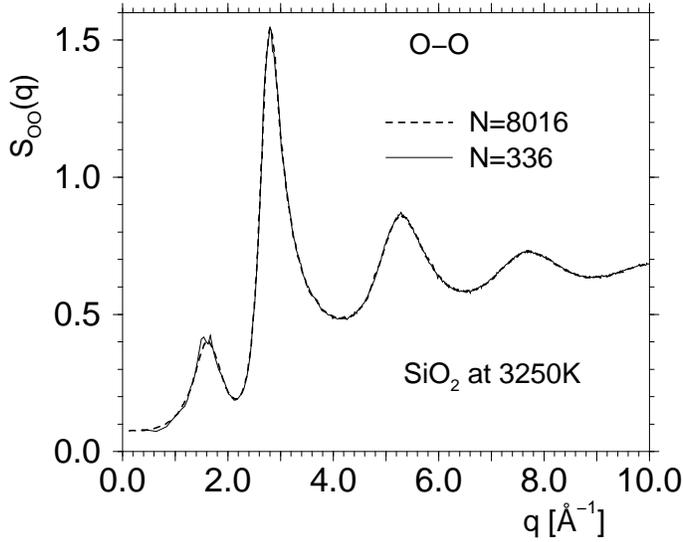}
\caption[]{
Partial structure factor for the O-O correlation in amorphous SiO$_2$ at
the density 2.36g/cm$^3$. The two curves correspond to two system sizes
(box size 16.8\AA~and 48.37\AA). It is clear that within the accuracy
of the data no difference is seen.}
\label{fig_structure_fin_size_sio2}
\end{figure}

\begin{figure}[hbtp]
\centering
\includegraphics[width=9.truecm]{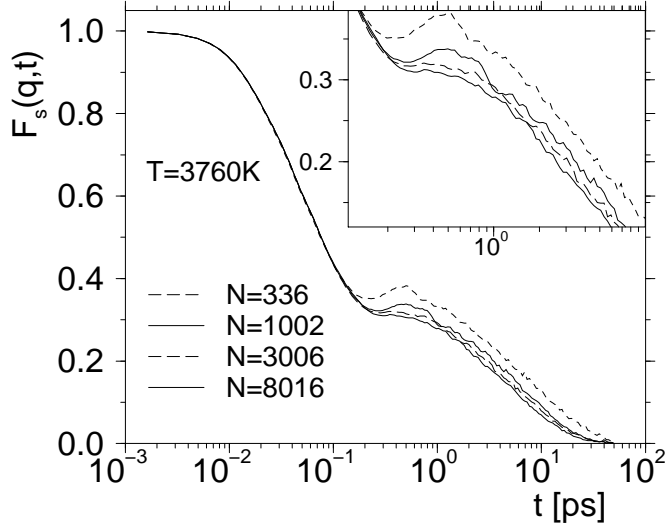}
\caption[]{
Time dependence of the intermediate scattering function of SiO$_2$
at 3760~K. The curves correspond to different system sizes and they demonstrate
that the relaxation dynamics in small systems is slower than the one in large
systems. Inset: Enlargement of the region close to the plateau.}
\label{fig_dynamics_fin_size_sio2}
\end{figure}

For the case of second order phase transitions it is well known
that the static as well as dynamic properties of the systems depend
strongly on the size of the system~\cite{stanley71,landau00}. For the
case of structural glasses such dependencies are much less pronounced,
although they are present. In Fig.~\ref{fig_structure_fin_size_sio2}
we show the partial static structure factor for the O-O correlation in
SiO$_2$~\cite{scheidler99}. The solid and dashed curves correspond to
the system sizes $N=336$ and $N=8016$ atoms and we see that within
the accuracy of our data there is no difference between the two
data sets. However, in other cases one does indeed find differences
in structural properties, e.g. in the distribution of the angles
between three neighboring particles in systems like Al-Ca-Si-O, if
the system size is below a few hundred atoms~\cite{ganster03}. Much
more sensitive to system size is the relaxation dynamics. In
Fig.~\ref{fig_dynamics_fin_size_sio2} we plot the intermediate
scattering function for the system sizes shown in the previous
figure. From this graph we recognize that the relaxation dynamics
is significantly slower than the one for the larger system and
that also the shape of the curves close to the plateau are not the
same~\cite{horbach96}. It has been shown that with decreasing system
size and decreasing temperature this difference in the dynamics
can become very pronounced in that for $N=100$ atoms the system
takes orders of magnitude longer to relax than the relaxation time
for large systems~\cite{horbach96,kim00,buchner99}. Although the
reason for this strong dependence is not understood very well it is
likely related to the fact that in small systems the connectivity
of the accessible configuration space is effectively smaller than in
larger systems, since in the latter fluctuations in density etc. can
be present that are strongly suppressed in the former. Note that
exactly the opposite is the case for the case of spin glasses:
There one finds that the smaller systems relax faster than the
large ones, at least if the interaction range is not too short
ranged~\cite{mackenzie83,brangian01,billoire01,brangian02b}.

\begin{figure}[hbtp]
\centering
\includegraphics[width=9.truecm]{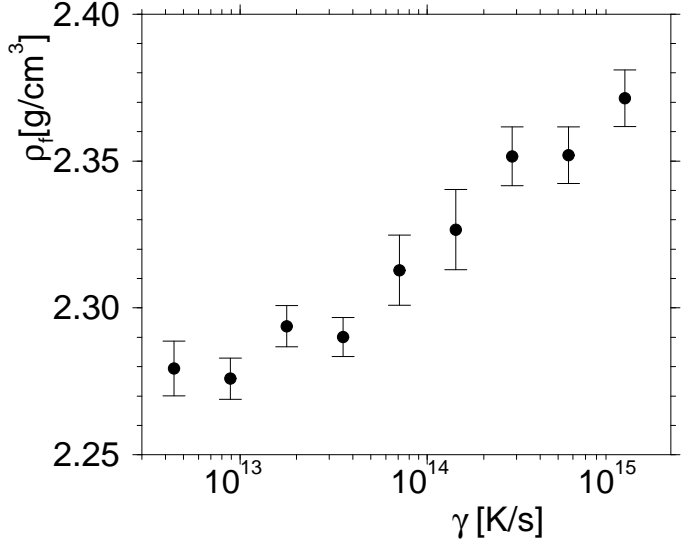}
\caption[]{
Density of SiO$_2$ at $T=0$ and zero pressure as a function of the cooling rate
with which the glass has been produced.}
\label{fig_cooling_rate_density}
\end{figure}

As already discussed in Sec.~\ref{sec2}, the glass transition is
basically just a kinetic phenomenon since it occurs at that temperature
at which the typical relaxation time of the system becomes comparable
with the time scale of the experiment (such as the inverse of the
cooling rate). This effect should not be forgotten if one uses
simulations to investigate the properties of glasses, since these
will in general depend quite substantially on the cooling rate. As an
example we show in Fig.~\ref{fig_cooling_rate_density} how in the case
of SiO$_2$ the density at low temperatures depends on the cooling rate
$\gamma$~\cite{vollmayr96}. (Note that these quenches have been done
at constant pressure.) From the figure we recognize that $\rho_f$ does
indeed depend significantly on $\gamma$ and that it is far from obvious
how this function should be extrapolated to the cooling rates that are
typically used in a real experiment (10K/s). The density is a quantity
that is relatively benign since it averages over the whole microscopic
structure. More microscopic structural quantities, such as the structure
factor, local coordination numbers, etc., show a even more pronounced
$\gamma-$dependence~\cite{vollmayr96}. Since in general it is not known
how the dependence of such quantities on $\gamma$ has to be extrapolated
to experimental cooling rates, a comparison with experimental data
becomes difficult or even impossible. (But of course many people sweep
all these methodological difficulties under the rug and go on to compare
happily their results with experimental data.)

\begin{figure}[hbtp]
\centering 
\includegraphics[width=9.truecm]{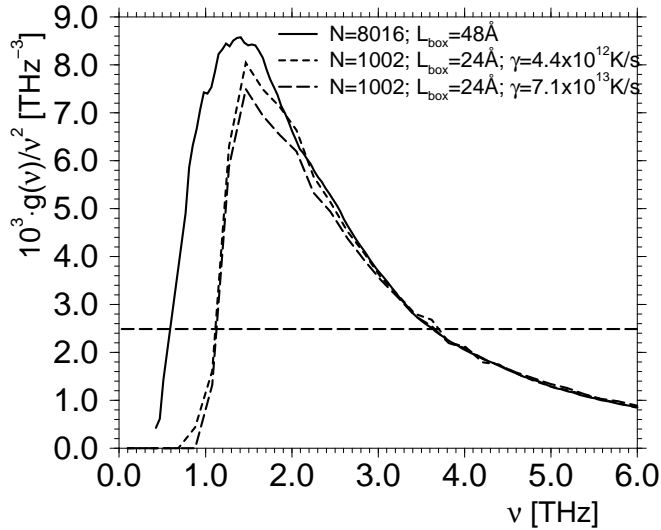}
\caption[]{
Density of state of amorphous silica divided by $\nu^2$ for two different 
system sizes and two
different cooling rates.}
\label{fig_cooling_rate_dos}
\end{figure}

Above we have commented on the observation that structural properties
of glasses are relatively insensitive to finite size effects once
the system is larger than a few hundred particles. There are, however,
situations were substantially larger systems are necessary. E.g. in recent
years a significant experimental and theoretical effort has been made to
understand the nature of the boson peak, mentioned in Sec.~\ref{sec2} (see
Fig.~\ref{fig_correlator_schematic})~\cite{benassi96,foret96,tara97a,dellanna98,schirmacher98,wischnewski98,rat99,sokol
ov99,goetze00,hehlen00,masciovecchio00,taraskin00a,grigera01,horbach01,foret02}.
Since this peak is an excess in $g(\nu)$, the vibrational density of
states, over the Debye-level, $g(\nu) \propto \nu^2$, it is reasonable
to plot the density of state divided by $\nu^2$ in order to see
whether or not an excess is seen. Note that in a simulation $g(\nu)$
can be calculated from the the eigenvalues of the dynamical matrix
or by the time Fourier transform of the velocity-auto-correlation
function~\cite{dove93}. In Fig.~\ref{fig_cooling_rate_dos} we show
$g(\nu)/\nu^2$ as a function of the frequency $\nu$ and we do indeed
see a peak at around $\nu\approx 1.5$THz~\cite{horbach99c}. However,
the plot also makes very clear that the location as well as the height
of the peak depends on the size of the simulation box as well as on the
cooling rate with which the glass has been produced. Also note that
for a sufficiently large system $g(\nu)/\nu^2$ should show at small
frequencies a plateau, the Debye-limit. The graph demonstrates however,
that even a system with more than 8000 ions is not sufficiently large
to show this Debye-law. Hence great care has to be taken if one starts
to analyse the modes and the mechanisms that give rise to the observed
peak, since it might very well be that one analyses modes that in a
realistically large system are irrelevant.

Before we end this section we briefly make some general comments
on the systems one simulates. Roughly speaking one can distinguish
two types of simulations: In the first type one tries to obtain some
general understanding of the properties of glass-forming systems that
subsequently can be used to check theoretical concepts or theories
(properties of the energy landscape, applicability of MCT, search for
diverging length scales, etc.). In such simulations one does not really
care whether or not the simulated system exists also in reality, as long
as the model does not have any pathological features. Therefore one often
uses very simple models such as lattice gases, Lennard-Jones fluids,
spin systems on a lattice, etc. This is in contrast to the situation
in which one uses a simulation to understand a particular feature of a
given material. In that case it is necessary to consider models that are
much more realistic since otherwise the feature of interest might even
be completely absent in the simulation (e.g. it is not possible to
study the process responsible for the light scattering mechanism by using
a simple Lennard-Jones model since there are no polarization effects).
The extreme limit of such a realistic calculation would of course be to
solve the Schr\"odinger equation for the many-body problem, but of course
it is clear that this is computationally not possible. The best thing one
presently can do is to make a so-called {\it ab initio} calculation in which
the particles obey Newton's equation of motion with a force field that is
calculated directly from the wave-function of the system~\cite{car85,marx00}. Such
calculations are presently feasible for systems of $O(100)$ particles and
time scales of 10ps. As we have seen above, these sizes and time scales
are for most cases insufficient and therefore one makes often the further
approximation to use {\it effective} force field that are just the sum of
a pairwise interaction potential between the particles. (Sometimes one
takes into account also three-body interactions.) These potentials will
depend on certain parameters (particle size, effective charges, etc.) that
are obtained by fitting the potential to the results from {\it ab initio}
calculations or to experimental data. Although the so obtained potentials
might in some cases be surprisingly accurate, they are an approximation
and thus it might happen that an effective potential that is reliable
for one quantity, is not very realistic for another one~\cite{benoit02}.
Therefore it is very advisable to check critically to what extend the
results that are predicted by a simulation with an effective potential
depend on this potential. An instructive example for this can be
found in Ref.~\cite{hemmati98}, reproduced as Fig.~1 in Ref.~\cite{kob99},
where the authors compare the temperature dependence of the diffusion
constant for different potential of SiO$_2$. They found that although
the structural properties of the various models are quite similar, the
relaxation dynamics differs by orders of magnitude! Thus this clearly
shows that it is not easy to use simulations to predict the reality.

\section{The Relaxation Dynamics of glass-forming Liquids as investigated
by Computer Simulations}
\label{sec5}

In the previous sections we have described some of the properties of
glass-forming liquids. In the following we will present the results
of computer simulations that have been used to investigate these
properties as well as to test to what extend the mode-coupling theory
of the glass transition is able to describe the relaxation dynamics of
these systems. In particular we will discuss two type of systems: A binary
Lennard-Jones mixture (BLJM) and silica (SiO$_2$). As we will see, these
two systems have very different structural and dynamical properties
and therefore it is of interest to see which dynamical features are
nevertheless common and can be rationalized by the theory.

Although in this section we will mainly discuss the simulations of
the BLJM system and silica, there is a multitude of results for other
glass-forming systems, such as hard spheres, soft sphere systems, water,
etc.~\cite{bernu87,miyagawa90,pastore88,roux89,barrat90b,signorini90,miyagawa91,laird91,sindzingre92,rino93,bala94,lewis94,sarnthein95,baschnagel96,guissani96,hurley96,mazzacurati96,kob97b,sciortino97b,guillot97,parisi97,badro98,bennemann98,foley98,kammerer98,kammerer98b,ribeiro98,sastry98,wilson98,zon98,bennemann99,berthier99,boeddeker99,benoit00,berthier00,doliwa00,gallo00,mossa00,aichele01,jund01b,lacks01,saika01,starr01,bagchi02,berthier02,nave02,puertas02,scheidler02,horbach02b,vollmayr02}.
It is found that many of the properties of these latter systems are
qualitatively similar to either the BLJM system or to silica. Thus
from this point of view we do indeed discuss the salient features of
glass-forming liquids and not only the properties of two particular
systems.

\subsection{Static and dynamic properties of a simple liquid with Newtonian
dynamics}
\label{sec5.1}

In this subsection we will discuss the properties of a simple
liquid in its normal and supercooled state. (Note that ``simple
liquid'' means that the interactions are short ranged and pairwise
additive~\cite{hansen86}.) Normally one-component simple liquids are
not good glass-formers, since they crystallize rapidly even if they are
supercooled only modestly. Therefore one usually considers binary systems
since the additional disorder introduced by the different species is
sufficient to prevent that the system crystallizes, at least on the time
scale of present days computer simulations (O($10^8$) time steps). One
system that seems to have very good glass-forming properties is a binary
mixture of Lennard-Jones particles with interactions that were chosen
to mimic the system Ni$_{80}$P$_{20}$~\cite{weber85}. In the BLJM
model we have a 80:20 mixture of particles, which in the following we
will call A and B particles, that interact via the potential

\begin{equation}
V_{\alpha\beta}(r)=4\epsilon_{\alpha\beta}
[(\sigma_{\alpha\beta}/r)^{12}-(\sigma_{\alpha\beta}/r)^6] \quad \mbox{with} \quad
\alpha, \beta \in \{{\rm A,B}\}
\label{eq36}
\end{equation}

\noindent
The parameters of the potential are given by $\sigma_{\rm AA}=1.0$,
$\epsilon_{\rm AB}=1.5$, $\sigma_{\rm AB}=0.8$, $\epsilon_{\rm
BB}=0.5$, and $\sigma_{\rm BB}=0.88$. Hence we use $\sigma_{\rm AA}$
and $\epsilon_{\rm AA}$ as the unit of length and mass, respectively
(setting the Boltzmann constant $k_B=1$). Time will be measured in units
of $(m\sigma_{AA}^2/48\epsilon_{AA})^{1/2}$, where $m$ is the mass of
the particles. The number of particles used was 1000 and the cubic
simulation box had the size $(9.4)^3$. The equations of motion were integrated
by means of the velocity-Verlet algorithm (Eq.~(\ref{eq33})) using a
time step of 0.01 and 0.02 at high ($T\geq 1.0$) and low temperatures
($T<1.0$), respectively.

\begin{figure}[hbtp]
\centering 
\includegraphics[width=9.truecm]{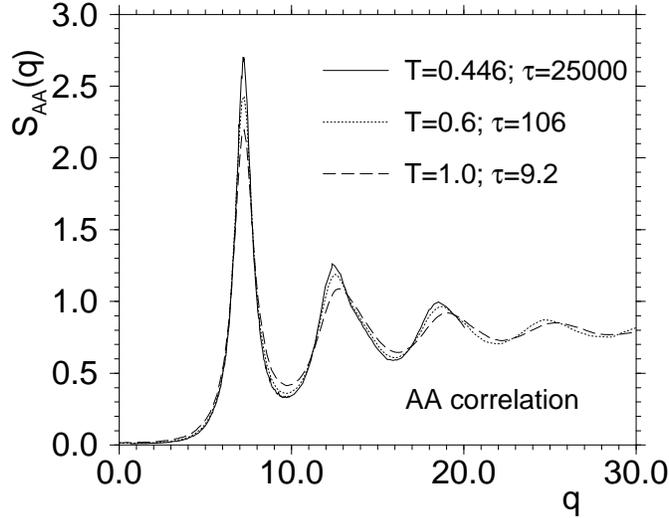}
\caption[]{
Wave-vector dependence of the partial structure factor $S_{\rm AA}(q)$
for a mixture of Lennard-Jones particles. The different curves correspond
to different temperatures. Also included are the $\alpha-$relaxation times $\tau$.}
\label{fig_structure_factor_lj}
\end{figure}

In Fig.~\ref{fig_structure_factor_lj} we show the wave-vector dependence
of the static structure factor for the A particles at three different
values of $T$~\cite{gleim_diss}. (Note that in a binary system there are three partial
structure factors~\cite{hansen86}.) From this graph we recognize that in
the $T-$range considered, the structure does not change significantly. The
only effect seen is that a decrease of $T$ gives rise to peaks and
minima that are more pronounced. (If the simulation would have been
done at constant pressure, also the location of the peaks would show
a weak $T-$dependence since the system expands or contracts if the
temperature is changed.) Since also the two other partial structure
factors do not show a more pronounced $T-$dependence, we conclude that
this quantity is indeed rather insensitive to a change in temperature,
in agreement with the comments on this in Sec.~\ref{sec2}.

Although the structural and thermodynamic properties of the system show
only a very mild $T-$dependence, the dynamic properties show a very
strong one. To demonstrate this we have included in the figure also the
$\alpha-$relaxation time $\tau(T)$ at the three different temperatures
(see below for the precise definition of $\tau(T)$). We see that in the
temperature range considered this time increases by a factor of more
than $10^3$, showing that the dynamic properties of the system do indeed
change much faster than the structural ones.

The relaxation time $\tau$ characterizes the dynamics only on the time
scale of the $\alpha-$relaxation, i.e. on the longest time scale of the
system. In order to understand the relaxation dynamics also on the other
time scales it is useful to consider time correlation functions. One
important example for such a correlation function is the mean-squared
displacement (MSD) of a tagged particle which is defined by

\begin{equation}
\langle r^2(t) \rangle = \frac{1}{N_\alpha} \sum_{i=1}^{N_\alpha} 
\langle |{\bf r}_i(t) - {\bf r}_i(0)|^2 \rangle   \quad .
\label{eq37}
\end{equation}

\begin{figure}[hbtp]
\centering
\includegraphics[width=9.truecm]{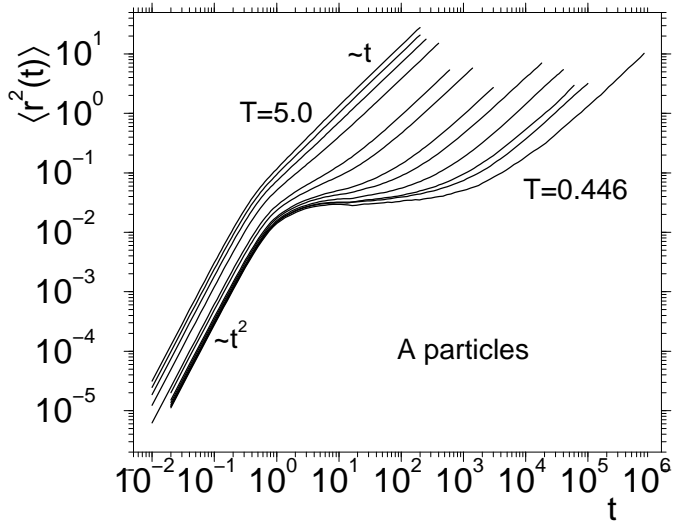}
\caption[]{
Time dependence of the mean-squared displacement of the A particles
for BLJM system. The curves correspond to different temperatures that
are given by $T=5.0$, 4.0, 3.0, 2.0, 1.0, 0.8, 0.6, 0.55, 0.50, 0.475,
0.466, and 0.446 (from left to right).}
\label{fig_msd_lj_nd}
\end{figure}

In Fig.~\ref{fig_msd_lj_nd} we show the time dependence of $\langle
r^2(t)\rangle$ for the A particles for different temperatures. Let
us start the discussion of the data with the curves for high $T$. For
short times we see that $\langle r^2(t) \rangle \propto t^2$, since we
are in the ballistic regime mentioned in Sec.~\ref{sec2}. Once $\langle
r^2(t)\rangle$ is of the order 0.04, i.e. the distance is around 0.2,
the time dependence changes over to a power-law with exponent 1.0,
i.e. the system is diffusive. This change in the $t-$dependence is due to
the collisions of the tagged particles with its neighbors that surround
it at $t=0$. Despite the presence of these neighbors the particle is
still able to move away quickly from its initial position, i.e. $\langle
r^2(t)\rangle$ increases quickly with time.

Also for low temperatures we see at early times a ballistic behavior. In
contrast to the curves for high $T$, the mean squared displacement
does at the end of this regime not cross over to the diffusive regime,
but instead shows a plateau at intermediate times. This plateau is due
to the cage effect mentioned in Sec.~\ref{sec2}, i.e. the temporary
trapping of the particle by its neighbors.  Only for sufficiently long
times the particle is able to leave this cage and once it has done so
the MSD quickly crosses over to a $t-$dependence that corresponds to a
diffusive motion. Note that the height of the plateau is around 0.03,
from which is follows that the size of the cage is around 0.17, i.e. it
is relatively small compared to the typical nearest neighbor distance
which is around 1.0~\cite{kob95a}.

The mean squared displacement gives the typical {\it average} distance
that a tagged particles moves within a time $t$. It is of course also
of interest to investigate the distribution of these distances. This can be
done by means of the self part of the van Hove function which is defined
as~\cite{hansen86,balucani94}:

\begin{equation}
G_s^{\alpha}(r,t) = \frac{1}{N_\alpha} \sum_{i=1}^{N_\alpha}
\langle \delta ( r-|{\bf r}_i(t)- {\bf r}_i(0)| )\rangle \quad .
\label{eq38}
\end{equation}

\begin{figure}[hbtp]
\centering
\includegraphics[width=6.5truecm]{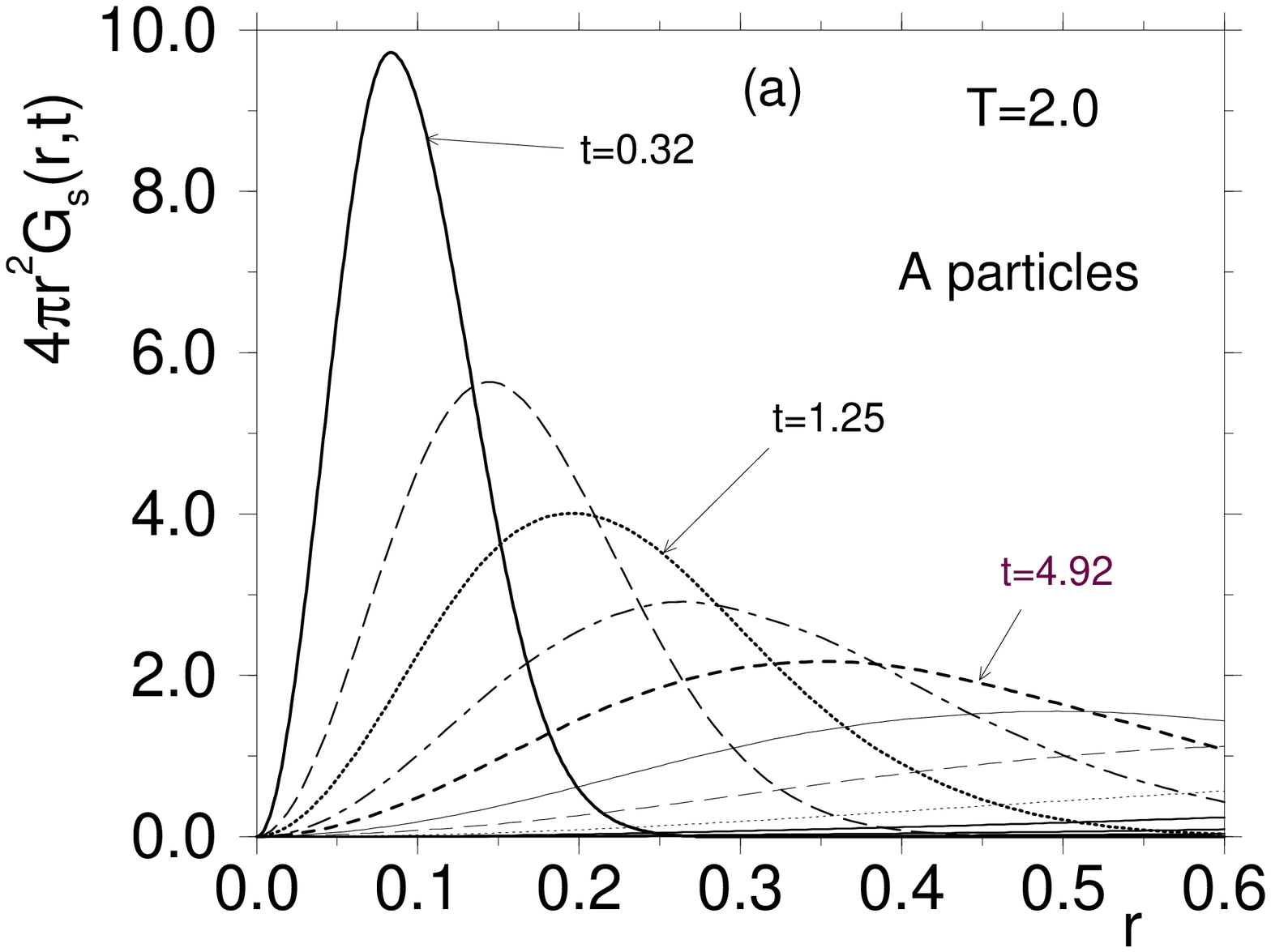}
\includegraphics[width=6.5truecm]{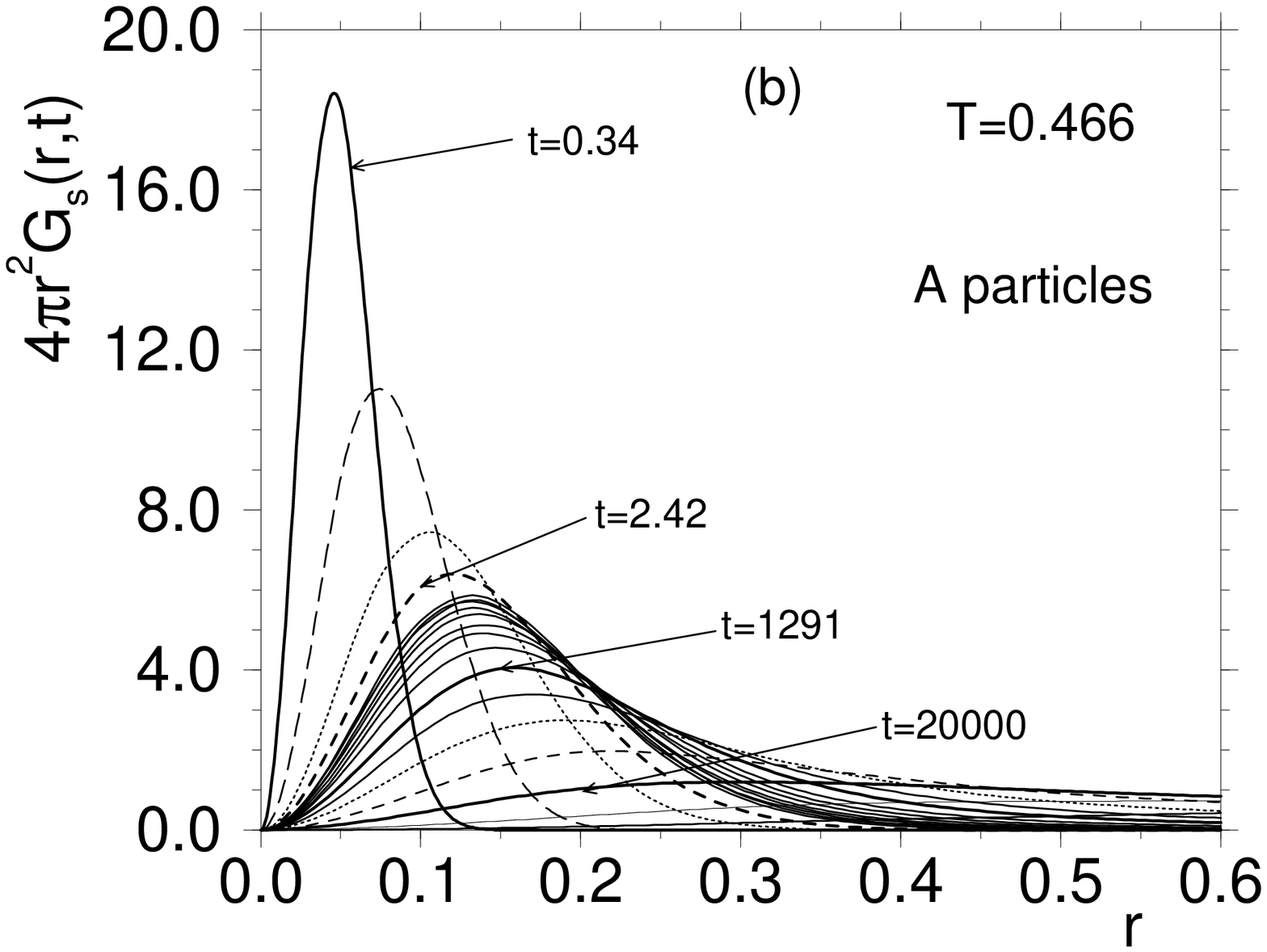}
\caption[]{
$r-$dependence of the van Hove function for the A particles in the BLJM at a
high and low temperature (panel (a) and (b), respectively). The
curves correspond to times $t=0.32\cdot 2^n$, $n=1, ...$.}
\label{fig_gs_vs_r_nd_lj}
\end{figure}

\noindent
Thus we see that $G_s^{\alpha} (r,t)$ is the probability that a particle
of type $\alpha$ has moved within the time $t$  exactly a distance $r$
and that $\langle r^2(t) \rangle$ is just the second moment of this
distribution. Furthermore we recognize that $G_s^{\alpha} (r,t)$
is nothing else the the space Fourier transform of the incoherent
intermediate scattering function defined in Eq.~(\ref{eq16b}). In
Fig.~\ref{fig_gs_vs_r_nd_lj} we show $4\pi r^2 G_s^{\rm A}(r,t)$ for
different values of $t$.
(The reason to plot $4\pi r^2 G_s^{\rm A}(r,t)$ instead of $G_s^{\rm
A}(r,t)$ is that it is the former that is measured in the simulation and
that will enter most theoretical formula, since the system is isotropic.)

From the graph that corresponds to high $T$ we see that at short times
the van Hove function is a single peak that moves quickly to the right
(its location moves $\propto t^2$). It is found that this peak is just
a Gaussian distribution multiplied with $4\pi r^2$. Also for times that
are longer than the duration of the ballistic regime, the curve is a
Gaussian. Only now the location of the peak moves proportional to $t$,
since the dynamics of the particles is diffusive.

For low temperatures and short times the dependence of $G_s^{\rm A}(r,t)$
on $t$ and $r$ is similar to the one at high $T$. Also at very long
times the function is again given by a Gaussian. For {\it intermediate}
times we see however a very different behavior in that the curves for
the lower temperature hardly depends on time. Thus we see here directly
the cage effect mentioned earlier, i.e. that the particles are trapped by
their neighbors. Note, however, that the curves still do depend on time
somewhat (although only weakly). A closer inspection of the data shows
that in this time window the {\it shape} of the curves does {\it not}
change and that their time dependence can be written as~\cite{kob95a}

\begin{equation}
G_s^{\alpha}(r,t) = G_s^{c,\alpha}(r) + H_{\alpha}(r) G(t) \quad ,
\label{eq39}
\end{equation}

\noindent
which is nothing else than a special case of the factorization property
shown in Eq.~(\ref{eq25}). Thus this is a first example that MCT is
able to rationalize a non-trivial behavior for this system. Similar
results have been found for the distinct part of the van Hove
function~\cite{kob95a}, i.e. the space-time correlation function in
which the second index $i$ in Eq.~(\ref{eq38}) is replaced by another
summation index $j$~\cite{hansen86,balucani94}.

\begin{figure}[hbtp]
\centering
\includegraphics[width=9truecm]{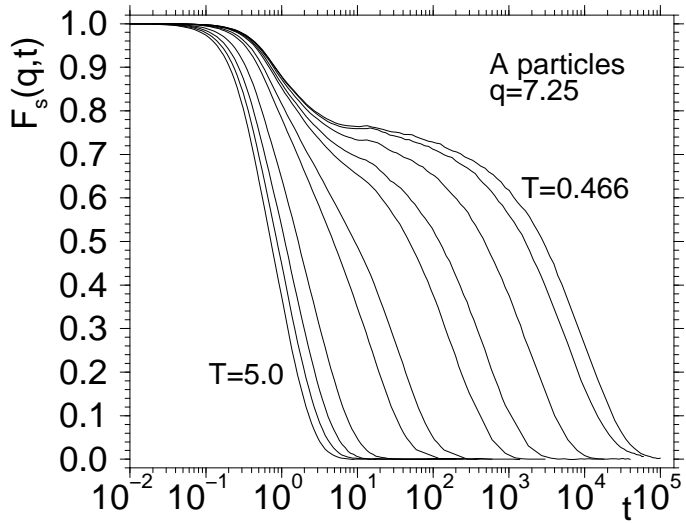}
\caption[]{
Time dependence of the incoherent intermediate scattering function for
different temperatures. The wave-vector corresponds to the location of the
main peak in the static structure factor.}
\label{fig_fs_vs_t_all_T_nd}
\end{figure}

The van Hove correlation function is a correlator that is very useful
to understand the dynamics of the particles in {\it real} space. However,
all scattering experiments are done in {\it reciprocal} space and hence it is
useful to investigate also how the intermediate scattering functions,
introduced in Eqs.~(\ref{eq16}) and (\ref{eq16b}), depend on temperature,
time and wave-vector. In Fig.~\ref{fig_fs_vs_t_all_T_nd} we show the
$t-$dependence of $F_s(q,t)$ for the A particles~\cite{kob95b}. In agreement with the
schematic drawing of Fig.~\ref{fig_correlator_schematic} we see
that at high temperature the correlator decays quickly to zero and a
closer inspection of the data shows that this decay is very close to the
expected exponential. At low temperature we see the two-step relaxation
discussed in the context of Fig.~\ref{fig_correlator_schematic}. From
this figure one clearly sees how strongly the relaxation dynamics
depends on temperature. Note that the curves do {\it not} show a dip
at intermediate times that in Fig.~\ref{fig_correlator_schematic}
was associated with the boson-peak.  The reason for this difference
is that liquids that have a structure in which the particles are very
closed packed do normally not have a boson-peak.  Instead it is a very
prominent feature in systems whose structure is given by an open network,
such as, e.g., silica. Finally we remark that the $T-$ and $t-$dependence
of $F_s(q,t)$ of the B particles as well as the one of the coherent
function $F(q,t)$ look qualitatively similar to the curves shown in
Fig.~\ref{fig_fs_vs_t_all_T_nd}~\cite{kob95b}.

\begin{figure}[hbtp]
\centering
\includegraphics[width=9truecm]{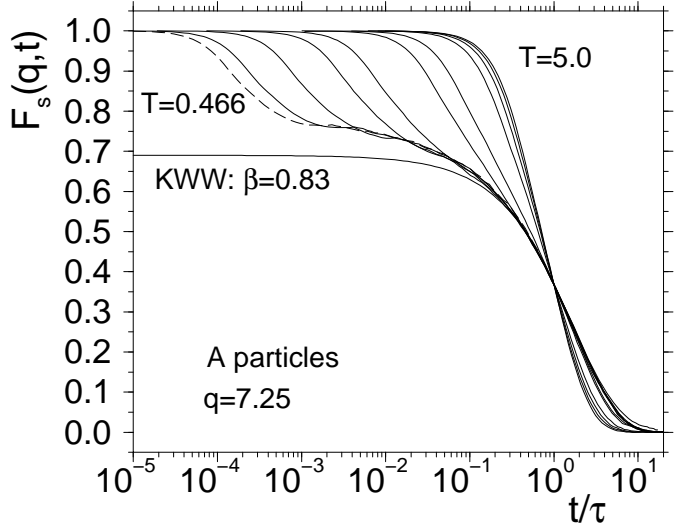}
\caption[]{
Incoherent intermediate scattering function as a function of rescaled time
$t/\tau(T)$ for the same correlators shown in Fig.~\ref{fig_fs_vs_t_all_T_nd}. Also
included is a fit to the curve at the lowest $T$ at long times with the KWW function
from Eq.~(\ref{eq3}).}
\label{fig_ttsp_lj}
\end{figure}

Using the time correlation functions just discussed, we can test
a further prediction of MCT, the validity of the time-temperature
superposition principle (see Eq.~(\ref{eq32})). For this we can define
an $\alpha-$relaxation time $\tau(T)$ by requiring that $F_s(q,\tau(T))=
e^{-1}$. In Fig.~\ref{fig_ttsp_lj} we show the correlators for the
different temperatures as a function of $t/\tau(T)$. We see that we have
basically two temperature regimes. At high temperatures the curves fall
at long times quite well onto a master curve that is described well by
an exponential. The curves at low $T$ also fall on top of each other,
but this time the master curve is a stretched exponential which is
included in the figure as well. (Note that the curve for $T=0.466$
(dashed curve) coincides so well with the KWW-function that the former
cannot be seen anymore at long rescaled times.) Similar master curves
are found for other correlation functions and hence we conclude that
the prediction of MCT regarding the validity of the TTSP holds for the
present system. We also mention that the stretching exponent $\beta$
is relatively large ($\beta=0.83$). Its value does, however, depend on
the correlator considered (different wave-vector or species) and for
the present system values as small as 0.5 have been found~\cite{kob95b}.

\begin{figure}[hbtp]
\centering
\includegraphics[width=9truecm]{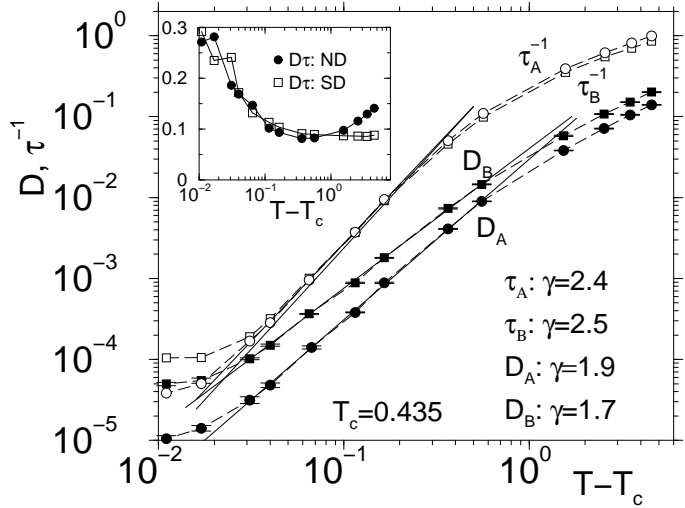}
\caption[]{
Main figure: Tagged particle diffusion constant $D$ and
$\alpha-$relaxation time as a function of $T-T_c$ for the BLJM system
(curves with symbols). The straight lines are fits to the data with the
power-law given by Eq.~(\protect\ref{eq24b}) and $\gamma$ is the exponent
resulting from these fits. The inset shows the product $D \tau$ vs $T-T_c$ for the ND
(filled symbols) and the SD discussed in Sec.~\ref{sec5.2} (open symbols).}
\label{fig_diffconst_tau_a_b_nd}
\end{figure}

From the mean-squared displacement one obtains immediately the diffusion
constant $D$ via the Einstein relation $D= \lim_{t\to \infty} \langle
r^2(t) \rangle /6t$. From the intermediate scattering function we have
determined the $\alpha-$relaxation time $\tau(T)$. Thus we now can check
whether the $T-$dependence of $D$ and $\tau$ is indeed given by the
power-law predicted by MCT (see Eq.~(\ref{eq24b})). That this is in fact
the case is demonstrated in Fig.~\ref{fig_diffconst_tau_a_b_nd} which is a
log-log plot of $D$ and $\tau^{-1}$ vs. $T-T_c$. We see that at low, but
not too low, temperatures the data does indeed fall on a straight line,
i.e. that we have a power-law dependence. Note that we have fixed the
value of $T_c$ for all four data sets to $T_c=0.435$, which shows that
such a power-law can be seen using the {\it same} critical temperature,
as predicted by the theory. We also note that the values of the exponents
$\gamma$ for the two relaxation times are very close to each other
(according to MCT they should be the same) but that they differ from
the one for the diffusion constant. According to Eq.~(\ref{eq24b}) the
product $D(T)\tau(T)$ should be a constant if one is close to $T_c$. In
the inset of Fig.~\ref{fig_diffconst_tau_a_b_nd} we show this product
(filled circles) and we recognize that it does depend on $T$, but that
this dependence is rather mild compared to the one of $D$ or $\tau$. (We
also note that if this product is calculated for the stochastic dynamics
discussed in Sec.~\ref{sec5.2}, it is basically independent of $T$
for high and intermediate temperatures, but shows at low $T$ the same
$T-$dependence as the curve for the ND, see the open symbols in the
inset.) The reason for this residual $T-$dependence is twofold: First
of all the result given by Eq.~(\ref{eq24b}) holds only very close to
$T_c$. A recent theoretical MCT calculation for a binary system has shown
that if one is not extremely close to $T_c$ the product is indeed not
constant and it is likely that one reads off from the $D(T)$ and $\tau(T)$
data only an effective exponent~\cite{voigtmann02}. The second reason
is that very close to $T_c$ it cannot be expected that the predictions
of MCT are valid, since the hopping processes mentioned at the end of
Sec.~\ref{sec3} will become important and modify the result given by
Eq.~(\ref{eq24b}). That such hopping processes are probably present at the
two lowest temperatures can be seen in Fig.~\ref{fig_diffconst_tau_a_b_nd}
in that the data points do no longer fall on the straight line for
the power-law.

Finally one wonders of course which one of the exponents $\gamma$
is the correct one. Although there is unfortunately no simple answer
to this question the general procedure to determine $T_c$ and $\gamma$
(and hence the other exponents $a$ and $b$) is as follows: The exponents
$\gamma$ can be determined from the $T-$dependence of the relaxation
times of various correlators as well as from the one of the diffusion
constant(s). The exponents $a$ and $b$ (or more general the exponent
parameter $\lambda$ given in Eq.~(\ref{eq27}) or (\ref{eq30})) can be
obtained from fitting various time correlation functions (at different
temperatures) in the $\beta-$regime with the $\beta-$correlator. Since
all these parameters are connected in a one-to-one way to each other,
one has to optimize all these fits in a global way. This approach usually
allows to obtain reliable values for $\lambda$ and $T_c$. For the present
BLJM system it is found that the exponent $\gamma$ should indeed be
close to 2.4, i.e. that the values of $\gamma$ as determined from $D$
are not reliable~\cite{kob95b,nauroth97,gleim98,gleim00}. The reason why
$D$ is more likely of not giving the correct exponent are the mentioned
hopping processes: It is sufficient that a few percent of the particles
hop (leave their cage) to increase the MSD considerably, i.e. these rare
events can change the value of $D$ substantially, whereas these type of
events affect the average relaxation time only weakly. Hence it must be
expected that very close to $T_c$ the diffusion constant is dominated
by these hopping processes that are not taken into account by the ideal
version of the theory and that therefore the expected $T-$dependence
is modified.

\begin{figure}[hbtp]
\centering
\includegraphics[width=7truecm]{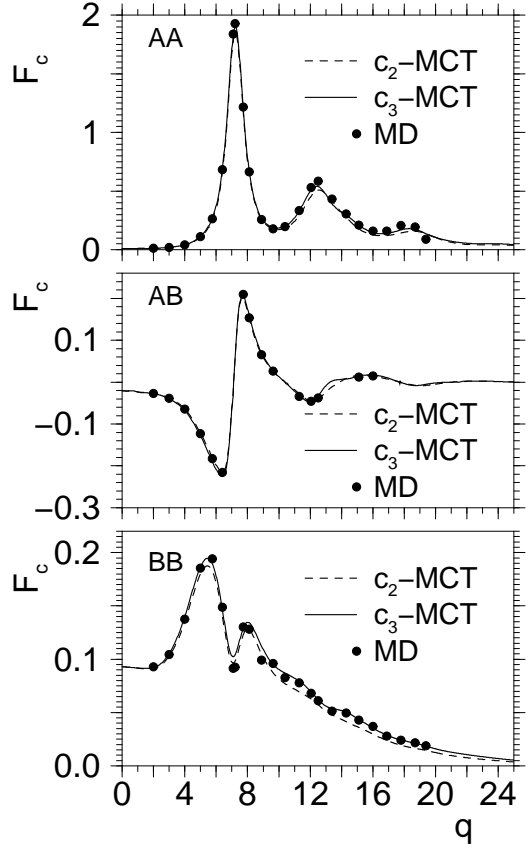}
\caption[]{
Wave-vector dependence of the non-ergodicity parameter for the BLJM
system. The dots are the results from the simulations and the dashed
and solid lines are the predictions of MCT neglecting and including the
$c_3-$terms, respectively (see text for details).}
\label{fig_nep_lj}
\end{figure}

All the results that we have discussed so far concern only {\it
qualitative} tests of MCT. However, in Sec.~\ref{sec3} we have mentioned
that in principle it is possible to use the theory also to do a
full microscopic calculation of the relaxation dynamics, if the static
quantities, such as the structure factor, are known. In the following
we will show that such a calculation is indeed possible, at least for
the present BLJM system. 

From the simulations described above, it is also possible to obtain the
wave-vector dependence of the three partial structure factors with high
accuracy, i.e. to within 1\%. From these $S_{\alpha\beta}(q)$ one can
obtain the direct correlation functions, $c_{\alpha\beta}(q)$, which in
turn can be used as input to calculate the vertices $V^{(2)}$ and hence
the memory function (see Eqs.~(\ref{eq19}) and (\ref{eq20})). Hence this
allows to obtain the prediction by MCT for the $q$ and $t-$ dependence
of the intermediate scattering function that can be compared directly
with the results of the simulation.

One quantity of interest for such a comparison is the non-ergodicity
parameter introduced in Eq.~(\ref{eq25}), i.e. the height of the
plateau in the $\beta-$regime. (If the correlator considered is
the coherent or incoherent intermediate scattering function, this
height is also known as the Debye-Waller and Lamb-M\"ossbauer-factor,
respectively.) In Fig.~\ref{fig_nep_lj} we show the $q-$dependence
of the non-ergodicity parameters for the three partial coherent
intermediate scattering functions. The symbols are the results from
the simulations~\cite{gleim98} and the solid and dashed lines are the
predictions of MCT. The dashed line corresponds to the version of
the theory in which the vertices $V^{(2)}$ in Eq.(\ref{eq25}) depend
only on the structure factor, i.e. the contribution to $V^{(2)}$ from
the three-particle correlations $c_3$ are neglected, whereas the
solid line corresponds to the case where $c_3$ has been taken into
account~\cite{nauroth97,sciortino01}. (Note that the function $c_3({\bf
q},{\bf k})$ has been determined also directly from the simulation.) From
Fig.~\ref{fig_nep_lj} we recognize that these two versions of the
theory give basically the same prediction for the $q-$dependence of the
non-ergodicity parameters and hence we can conclude that for this system
the effect of $c_3$ is very small, in agreement with the results for a
system of soft spheres~\cite{barrat89}.

From the figure we also recognize that the agreement between the
theoretical curve from the $c_3-$MCT and the data from the simulation
is excellent in that also small details are reproduced very well. We
emphasize that {\it no} fit parameter of any kind has been used to
calculate these theoretical curves.  Hence we conclude that the theory
is indeed able to predict highly non-trivial quantities of the relaxation
dynamics with very good accuracy.

Apart from the $q-$dependence of the non-ergodicity parameters it is also
quite simple to calculate the value of the critical exponent $\gamma$,
which the theory predicts to be 2.34~\cite{nauroth97}, in very good
agreement with the values 2.4-2.5 found from the simulations. Also the
$q-$dependence of the amplitudes $h$ occurring in Eq.~(\ref{eq25}) for the
coherent and incoherent scattering function is predicted reliably. The
only quantity for which a significant discrepancy between theoretical
prediction and simulation is found is the critical temperature $T_c$ for
which MCT predicts a value of 0.92 whereas the value from the simulation
is 0.435~\cite{nauroth97}. Why there is such a large difference and even
more puzzling why despite this large difference all the other predictions
of the theory are so accurate, is presently not well understood. It has
to be mentioned, however, that this observation does not only hold for
the present BLJM system but has also been found for the case of hard
spheres and of soft spheres~\cite{barrat89,megen94,gotze99}.

\begin{figure}[hbtp]
\centering
\includegraphics[width=9truecm]{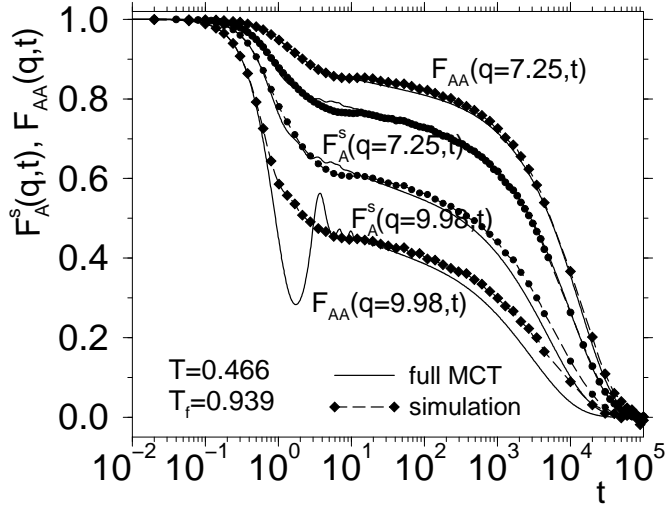}
\caption[]{
Time dependence of the coherent and incoherent intermediate scattering
function for the A particles. The symbols are the results of the
simulation and the solid lines the prediction of MCT.The two wave-vectors
correspond to the location of the first maximum and minimum in the static
structure factor.}
\label{fig_fs_vs_t_nd_mct}
\end{figure}

The quantitative tests of the MCT that we have presented so far concerned
the dynamics on intermediate and long times, i.e. the time scale of
the $\beta-$ and $\alpha-$relaxation. As we have already mentioned
in Sec.~\ref{sec3}, this dynamics is predicted to be independent of
the microscopic dynamics (and below we will give evidence that this is
indeed the case). However, if one wants to use the theory to predict
the full time dependence of the correlation functions it is necessary
to have also a good theory for the relaxation dynamics at {\it short}
times, i.e. to know the memory function $M^{\rm reg}$ in Eq.~(\ref{eq17}).
As already mentioned in Sec.~\ref{sec3} such a theory does presently not yet
exist and therefore one has to content oneself with a phenomenological Ansatz
for $M^{\rm reg}$. One model that has been found to work well is given by

\begin{equation}
M^{\rm reg} = \alpha(q)/\cosh(\beta(q)t) \quad,
\label{eq40}
\end{equation}

\noindent
where $\alpha(q)$ and $\beta(q)$ are constants that
can be determined directly from sum rules over static
quantities~\cite{tankeshwar87,tankeshwar95,hansen86,boon80}. Using this
functional form of $M^{\rm reg}$, the time and $q-$dependence of the solution
of the MCT equations can be obtained numerically. The temperature
dependence of these solutions enters only via the $T-$dependence of
the static structure factor and trivial kinetic factors. However,
as we have mentioned above, the theory overestimates the value of the
critical temperature $T_c$ at which the system freezes ($T_c^{\rm sim}=
0.435$ and $T_c^{\rm MCT}=0.92$).  This means that close to $T_c^{\rm
sim}$ the value of the vertices in the memory function are already too
large. To fix this problem we have evaluated this memory function at a
temperature $T_f(T)$ and have adjusted $T_f$ such that the time scale for
the $\alpha-$relaxation as predicted by the theory for $F_{\rm AA}(q,t)$
at $q=7.25$ matched the one of the simulation at temperature $T$. More
details on this can be found in Refs.~\cite{nauroth99,kob02}.

In Fig.~\ref{fig_fs_vs_t_nd_mct} we compare the time dependence
of the coherent and incoherent intermediate scattering function as
obtained from the simulation~\cite{kob95b,kob94} and as predicted by
the theory~\cite{nauroth99,kob02}. The two wave-vectors correspond
to the location of the minimum and maximum in the partial structure
factor for the A particles.  From this figure we recognize that at
intermediate and long times the agreement between theory and simulation
is very good in that the height of the plateau, the time scale for the
$\alpha-$relaxation, as well as the shape of the $\alpha-$relaxation
are very similar. The main discrepancies are found for $F_{\rm AA}(q,t)$
for the larger wave-vector in that at short times the theoretical curve
shows a strong oscillation that is not present in the data from the
simulation. This difference indicates that the effective damping of the
correlator at short times is not described correctly, i.e. that for this
value of $q$ the memory function $M^{\rm reg}$ is too small. We emphasize,
however, that this error has nothing to do with the inadequacy of the MCT
but instead is related to our insufficient understanding of the dynamics
of liquids at short times (see also Ref.~\cite{casas00}). In contrast
to this the relaxation dynamics of the system at intermediate and long
times is described very well by the theory since similar results as the
ones shown in Fig.~\ref{fig_fs_vs_t_nd_mct} are found also for the other
partial correlation functions and other wave-vector~\cite{nauroth99}.

\subsection{The relaxation dynamics of a simple liquid with stochastic
dynamics}
\label{sec5.2}

The results discussed in Section~\ref{sec5.1} concerned the static
and dynamical properties of a simple liquid with {\it Newtonian}
dynamics. However, in quite a few practical applications, such as
colloids or polymers in a solvent, the microscopic dynamics is better
described by a Brownian one. The goal of this subsection is therefore
to investigate to what extend the microscopic dynamics influences the
relaxation dynamics, and hence the slowing down of the dynamics upon
cooling, in a glass-forming system.

As we have seen in the previous sections, according to mode-coupling
theory a small change in structure can have a large effect on the
relaxation dynamics of the system, a result that is compatible with the
curves and relaxation times shown in Fig.~\ref{fig_structure_factor_lj}.
Hence if we want to study the influence of the microscopic dynamics
upon the relaxation dynamics it is imperative not to change the static
properties of the system. One possibility to do this is to use exactly
the same interaction potential that we have used to study the dynamic
properties of the BLJM system and to change only the equations of motion.
One alternative to Newton's equations of motion is a Brownian dynamics in
which the inertia term is neglected and instead one has a friction term
as well as an external noise. However, the numerical integrators that
exist for such types of equations are not very accurate and thus require
a small step size~\cite{ermak75,paul95}, which is of course very unpleasant for
systems in which the dynamics is slow. One possibility to circumvent this
problem is to use a ``stochastic dynamics'' (SD), in which the equations
of motion are still second order, but they include also a damping and
noise term~\cite{paul95,gleim98}:

\begin{equation}
m\ddot{{\bf r}}_i+\nabla_i \sum_l V_{il}(|{\bf r}_l-{\bf r}_i|)
=-\zeta \dot{{\bf r}}_i + {\bf \eta}_i(t)\quad.
\label{eq41}
\end{equation}

\noindent
Here $V_{il}$ is the potential between particles $i$ and $l$, ${\bf
\eta}_i(t)$ are Gaussian distributed random variables with zero mean,
i.e., $\langle {\bf \eta}_i(t)\rangle =0$, and $\zeta$ is a damping
constant. The fluctuation dissipation theorem relates $\zeta$ to the
second moment of ${\bf \eta}_i$, and we have $\langle {\bf \eta}_i(t)\cdot
{\bf \eta}_l(t') \rangle = 6k_B T \zeta \delta(t-t')\delta_{il}$. For
the following we will use a value of $\zeta$ of 10, which is large enough
that the results presented do not depend on $\zeta$ anymore (apart from
a trivial shift in the time scale). More details on this simulation can
be found in~\cite{gleim00,gleim_diss}. Note that all the {\it static}
properties of the system are completely independent of the value of
$\zeta$ and that $\zeta=0$ corresponds to the case of the Newtonian
dynamics.

\begin{figure}[hbtp]
\centering
\includegraphics[width=9truecm]{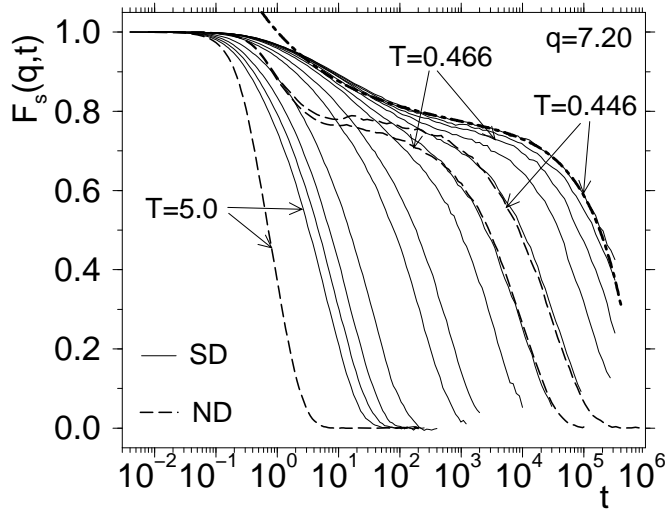}
\caption[]{
Time dependence of the incoherent intermediate scattering function for
the A particles. The solid lines are for the case of the stochastic
dynamics at the temperatures $T= 5.0$, 4.0, 3.0, 2.0, 1.0, 0.8, 0.6, 0.55, 0.5,
0.475, 0.466, 0.452, and 0.446. and the dashed lines are for the
Newtonian dynamics at $T=5.0$, 0.466, and 0.446. The dashed-dotted lines is a fit to
the SD curve at $T=0.446$ with the $\beta-$correlator from the MCT.}
\label{fig_fs_sd_nd_all_T}
\end{figure}

In Fig.~\ref{fig_fs_sd_nd_all_T} we show the time dependence
of the intermediate scattering function as obtained from the SD
(solid lines). The wave-vector is 7.20, i.e. is the location of
the maximum in the static structure factor. Although at a first
glance the set of curves look quite similar to the ones found for
the Newtonian dynamics (ND), see Fig.~\ref{fig_fs_vs_t_all_T_nd},
there are important differences. To simplify their discussion we have
included in the figure also the data from the latter dynamics at
certain temperatures (dashed lines). Comparing the corresponding data
at the highest temperature, we see that the curve for the ND decays
quicker to zero than the one for the SD and that the relaxation time,
which can, e.g., be defined again via $F_s(q,\tau)=e^{-1}$, is about 7
times shorter.  For the lowest temperature considered, this ratio has
grown to about 30. At these temperatures all curves show the plateau of
the $\beta-$relaxation. However, the correlators for the ND approach
this plateau very quickly whereas the ones for the D approach it very
smoothly. Thus we conclude that the way the particles explore the cage
is very different. Despite from these differences there are, however,
also quite a few similarities in the correlators from the two types of
dynamics: First of all the height of the plateau seems to be the same.
This means that the size of the cage is independent of the microscopic
dynamics (see also Ref.~\cite{lowen91}). Furthermore also the shape
of the correlator in the $\alpha-$regime is independent of the dynamics
as can be concluded from the fact that for certain temperatures the
dashed and solid curves trace each other very well and the fact that at low $T$
this shape is independent of $T$ (see Fig.~\ref{fig_ttsp_lj}). These
results are in agreement with the prediction of MCT since the theory
does indeed predict that the relaxation dynamics at intermediate and long
times is independent of the microscopic dynamics that is characterized
by the memory function $M^{\rm reg}(q,t)$ in Eq.~(\ref{eq17}).

Also included in the figure is a fit to the SD curve at the lowest
temperature with the $\beta-$correlator given by Eq.~(\ref{eq26})
(dashed-dotted curve). The value of $\lambda$ was fixed to 0.708, the
value that MCT predicts for this system~\cite{nauroth97} and is hence {\it
not} a fit parameter. (We mention that in order to get reliable fits it
was necessary to include also the correction to the $\beta-$correlator,
which according to MCT are of the form $t^{2b}$~\cite{gotze89}
(see also Refs.~\cite{gleim_diss,sciortino96,sciortino97} for more
details.) The figure shows that this functional form gives a very good
fit to the data for more than three decades in time, which is
strong evidence that MCT does indeed give a reliable description of
the $\beta-$relaxation.

\begin{figure}[hbtp]
\centering
\includegraphics[width=9truecm]{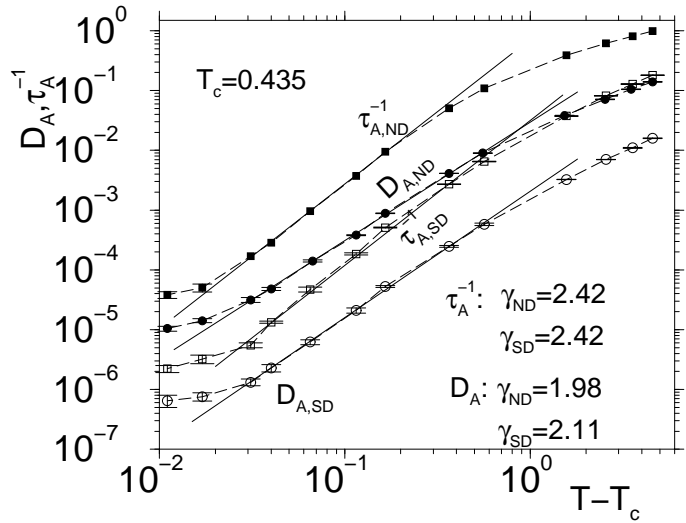}
\caption[]{
Temperature dependence of the relaxation time $\tau$ as determined
from the incoherent intermediate scattering function and the diffusion
constant (in both cases for the A particles). The filled and open symbols
correspond to the Newtonian and stochastic dynamics, respectively. The straight
lines are fits with a power-law using $T_c=0.435$ as the critical temperature. From
Ref.~\cite{gleim_diss}.}
\label{fig_d_tau_nd_sd}
\end{figure}

From the intermediate scattering function and the mean-squared
displacement of a tagged particle one can now determine again the
relaxation time $\tau(q,T)$ and a diffusion constant. According to
MCT the $T$ dependence of these quantities should be the same as their
$T-$dependence for the case of the ND. That this is indeed the case is
shown in Fig.~\ref{fig_d_tau_nd_sd} were we plot $D$ and $\tau^{-1}$
for the two types of dynamics. The value of the critical temperature $T_c$
has again been fixed to $T_c=0.435$, the value we have found for the
ND. The plot shows that with this value of $T_c$ also the data for the SD
are rectified for 2-3 decades in $D$ (or $\tau^{-1}$). Hence we conclude
that the power-law given by Eq.~(\ref{eq24b}) is found also for the
SD. Moreover, the exponents $\gamma$, given by the slope of the straight
lines, seems basically to be independent of the microscopic dynamics. In
particular the values of $\gamma$ as determined from $\tau(T)$ are the
same within the accuracy of the data. Hence we conclude from this figure
and the previous one that the whole $\alpha-$process is independent of the
microscopic dynamics, {\it apart from an overall shift in the time scale},
a result that is rather surprising.  From Fig.~\ref{fig_fs_sd_nd_all_T}
we recognize that this shift is only due to the different time dependence
of the dynamics at {\it short} times, i.e. by the different choice of
$M^{\rm reg}(q,t)$. Since we have shown in Fig.~\ref{fig_fs_vs_t_nd_mct}
that MCT is able to give a correct description of the relaxation dynamics
at long times, we hence conclude that if we would have a reliable theory
for the short time dynamics, we would be able to describe the relaxation
dynamics in the whole time range and for all microscopic dynamics. (Of
course this has to be restricted to the case that the microscopic
dynamics is ``local'', i.e. we do not allow, e.g., {\it global} Monte Carlo
moves~\cite{frenkel96,binder96,landau00}.)

\begin{figure}[hbtp]
\centering
\includegraphics[width=9truecm]{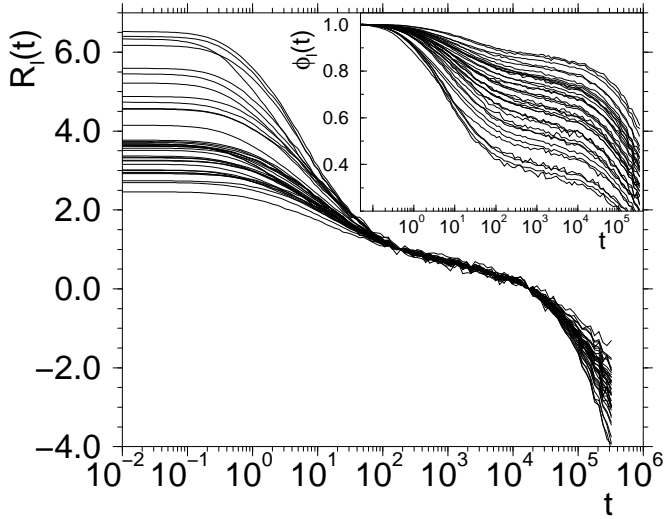}
\caption[]{
Main figure: Time dependence of $R_l(t)$ as defined in Eq.~(\protect\ref{eq42})
using various correlators (see main text for details). The time
dependence of these correlators are shown in the inset.}
\label{fig_factorization_sd}
\end{figure}

The last feature of the relaxation dynamics that we want to discuss is
the factorization property given in Eq.~(\ref{eq25}). Recall that this
equation says that on the time scale of the $\beta-$relaxation the shape
of all time correlation function is the same. In order to check whether
this is indeed true one could make fits to each of the correlation
functions with the $\beta-$correlator {\it using the same value of
$\lambda$}, as it has been done in Fig.~\ref{fig_fs_sd_nd_all_T}. Since
this procedure is quite involved it is, however, more advisable to use
an approach that has been proposed some time ago by Signorini {\it et
al.}~\cite{signorini90} and which goes as follows: Let $\Phi_l(t)$ be
an arbitrary correlator. From Eq.~(\ref{eq25}) it thus follows that the
following ratio is independent of $l$:

\begin{equation}
R_l(t) = \frac{\Phi_l(t)-\Phi_l(t')}{\Phi_l(t'')-\Phi_l(t')} \quad .
\label{eq42}
\end{equation}

Here $t'$ and $t''$ are two arbitrary times in the $\beta-$regime. If
Eq.~(\ref{eq25}) does indeed hold, the time dependence of $R_l(t)$
should just be the function $G(t)$, hence be independent of $l$. In
Fig.~\ref{fig_factorization_sd} we show the time dependence of
$R_l(t)$ for a total of 36 correlators, using times $t'=200$ and
$t''=15000$~\cite{gleim00}. These correlators include the coherent and
incoherent scattering function for the A and B particles at different
wave-vectors and their time dependence is shown in the inset of the
figure. From this inset we recognize that the time dependence of all
these correlators differs strongly in that the height of the plateaus,
the stretching etc. covers a wide range of values.  From the main figure
we see, however, that on the time scale of the $\beta-$relaxation all
these curves are indeed very similar in that the rescaling given by
Eq.~(\ref{eq42}) makes them to collapse nicely onto a master curve, the
shape of which is the function $G(t)$. Hence we can conclude that for
this system the factorization property predicted by MCT does indeed hold.

\subsection{Static and dynamic properties of a network forming liquid}
\label{sec5.3}

In the previous two subsections we have discussed the static and
dynamic properties of a {\it simple} liquid and have shown that this dynamics
can be rationalized very well with the help of MCT. The structure of
these simple liquids is quite close to the one of a close packing of
hard spheres. Already in Sec.~\ref{sec2} we have mentioned, however,
that the so-called strong glass-formers have a structure that is often
quite open and that the $t-$dependence of their relaxation dynamics is,
by definition, close to an Arrhenius law. Hence it is legitimate to ask
whether MCT is a useful theory for these type of systems as well. The
goal of this section is therefore to discuss the results of some computer
simulations that have been done to address this question.

In order to make a computer simulation of a strong glass-former it is
of course necessary to have a interaction model that has indeed the
required properties, i.e. an Arrhenius dependence of the relaxation
times. Fortunately silica, the paradigm of a strong glass-former, is an
important material not only for fundamental science but also in technical
applications as well as fields like geology.  Therefore there are a
multitude of potentials available that claim to give a good description
of this material~\cite{angell81,poole95}. One of the potentials that does
indeed seem to be quite reliable has been proposed some time ago by van
Beest, Kramer, and van Santen (BKS)~\cite{beest90}, and its functional
form is given by

\begin{equation}
\phi_{\alpha \beta}(r)=
\frac{q_{\alpha} q_{\beta} e^2}{r} +
A_{\alpha \beta} \exp\left(-B_{\alpha \beta}r\right) -
\frac{C_{\alpha \beta}}{r^6}\quad \alpha, \beta \in
[{\rm Si}, {\rm O}],
\label{eq43}
\end{equation}

\noindent
where $r$ is the distance between the ions of type $\alpha$ and $\beta$.
The values of the constants  $q_{\alpha}, q_{\beta}, A_{\alpha
\beta}, B_{\alpha \beta}$, and $C_{\alpha \beta}$ can be found in
Ref.~\cite{beest90}. Due to the presence of the Coulomb interactions
it is necessary to use an Ewald summation technique to calculate the
forces~\cite{frenkel96,binder96}, or one of the equivalent methods
proposed more recently~\cite{greengard87,solvason95,deserno98}. Therefore
making a simulation of these type of systems are presently about 10
times more expensive in computer time that models with only short range
interactions. As mentioned in Sec.~\ref{sec4} it is sometimes necessary
to use relatively large systems in order to avoid finite size effects
in the dynamics~\cite{horbach96,horbach01}. Therefore the following
results have been obtained for a system size of about 8000 atoms, which
corresponds to a box size of around 50\AA. More details of the simulation
can be found in Refs.~\cite{horbach99,horbach01b}.

\begin{figure}[hbtp]
\centering
\includegraphics[width=9truecm]{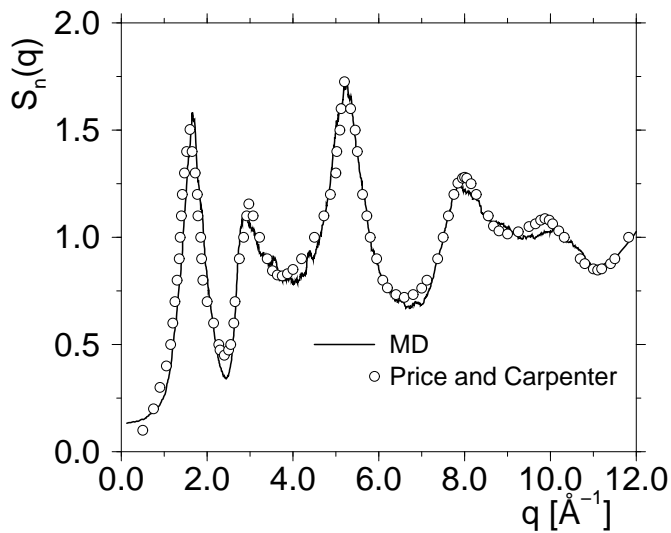}
\caption[]{
Wave-vector dependence of the structure factor of SiO$_2$ as measured in a
neutron scattering experiment. The solid line is the result of the simulation
with the BKS potential and the symbols are the experimental results by
Price and Carpenter~\cite{price87}.}
\label{fig_sio2_neutron_sq}
\end{figure}

Since we are modeling silica, e.g. a material that really exists in
nature, it is possible to compare the results from the simulation with
real experimental data (and in fact this should always be done in order
to check whether the potential used is indeed reliable, i.e. compatible
with known experimental facts). One such comparison is shown in
Fig.~\ref{fig_sio2_neutron_sq}, were we compare $S_n(q)$, the static
structure factor as measured in a neutron scattering experiment, from
the simulation with real experimental data. This function can be easily
calculated from the three partial structure factors $S_{\alpha\beta}(q)$
using the relation

\begin{equation}
S_n(q)=\frac{1}{N_{\rm Si}b_{\rm Si}^2+N_{\rm O}b_{\rm O}^2}
\sum_{\alpha,\beta} b_\alpha b_\beta S_{\alpha\beta}(q) \quad ,
\label{eq44}
\end{equation}

\noindent
where $b_\alpha$ are the neutron scattering length whose experimental
values can be found in the literature~\cite{susman91}. From the figure we
see that the agreement between the experiment and the simulation is good,
which shows that the BKS potential is quite reliable. \footnote{A brief
comment on the meaning of the various peaks: In a simple liquid, see e.g.
Fig.~\ref{fig_structure_factor_lj}, the first main peak corresponds to
the length scale of a nearest neighbor. However, for ionic systems like
silica this identification is not necessarily true, since the charged
atoms will induce a local ordering of the system. In the case of SiO$_2$
this is done in the form of tetrahedra, i.e. a silicon atom is surrounded
by four oxygen atoms. These tetrahedra are connected via their corners and
form an irregular open network. Thus in this system the first peak at in
the structure factor (located at 1.6\AA$^{-1}$) does not correspond to the
nearest neighbor distance (which would be a Si-O pair) but to the distance
between two neighboring tetrahedra. For even more complex systems, like
Na$_2$O-SiO$_2$, one can have structural features at even larger length
scales~\cite{horbach99b,horbach01c,jund01,horbach01,horbach02,meyer02}.} Hence we
can have some confidence that the dynamical properties as predicted by
the model will not be too far off from the ones of real silica.

\begin{figure}[hbtp]
\centering
\includegraphics[width=9truecm]{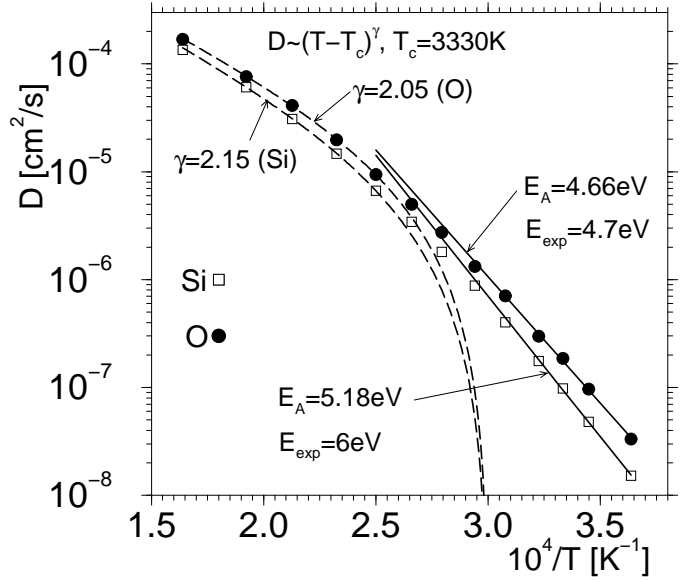}
\caption[]{
Arrhenius plot of the diffusion constant for the silicon and oxygen atoms as
predicted by the BKS potential. The solid straight lines at low $T$ are fits with
an Arrhenius law. The dashed lines at intermediate temperatures are fits with the
power-law given by Eq.~(\protect\ref{eq24b}).}
\label{fig_sio2_diffus}
\end{figure}

In Fig.~\ref{fig_sio2_diffus} we show the $T-$dependence of the
diffusion constants for the Si and O atoms (which were determined using
the Einstein relation)~\cite{horbach99}. Note that the temperature scale
covered is 6100~K~$\leq T \leq 2750$~K, i.e. the temperatures are rather
high. The reason for this is that silica is a very viscous system even
at high temperatures (the relaxation time at 2750~K is around 10ns, see
also Fig.~\ref{fig_angell}) and hence it is presently not possible to
equilibrate the system in a simulation at significantly lower temperature.
From the figure we recognize that at low temperatures the diffusion
constants show an Arrhenius dependence, as can be expected for this system
(see bold solid lines). The activation energies, given in the figure,
agree well with the experimental values~\cite{mikkelsen84,brebec80}. Hence we
conclude that the BKS potential is indeed able to give also a satisfactory
description of the relaxation dynamics of real silica.

What is surprising in the data from the simulation is that at high
temperatures there is a crossover from the Arrhenius dependence of $D$
to a weaker one, i.e.  there is a significant bend in the curves. (Note
that the viscosity and the relaxation times determined from the
intermediate scattering functions shows a very similar $T-$dependence,
hence this non-Arrhenius behavior is not a particularity of the diffusion
constant.) Such a $T-$dependence is reminiscent to the behavior found
in fragile liquids close to the critical temperature of MCT. Therefore
it is reasonable to check whether the power-law predicted by the theory,
see Eq.~(\ref{eq24b}), can also in this case be used to rationalize the
data. That this is indeed possible is demonstrated by the two dashed lines
shown in the figure that represent such a power-law dependence. If one
assumes that the value of $T_c$ is the same for Si and O, the resulting
critical exponents $\gamma$ are indeed very close together, in agreement
with the prediction of the theory. Also the $\alpha-$relaxation times
$\tau(T)$ show a power-law dependence and the critical temperature is
the same as the one found for the diffusion constant. However, as it was
already the case for the BLJM system, the value of the critical exponent
$\gamma$ as determined from $\tau$ is larger than the one found for $D$
($\gamma_\tau= 2.45$)~\cite{horbach01b}.

The critical temperature $T_c$ is found to be around 3330~K, a value
that is far above the melting temperature of the system ($T_m \approx
2000$~K). Thus we see that the critical temperature of MCT has nothing
to do with the system being in a supercooled state. Instead it is just
a temperature at which the transport mechanism of the atoms changes, as
already discussed in Sec.~\ref{sec3}. Last not least we mention that Hess
{\it et al.} have used experimental data for the viscosity to estimate
$T_c$ and have predicted a value around 3200~K~\cite{hess96,rossler98},
in surprisingly good agreement with the estimate from the simulation.

\begin{figure}
\centering
\includegraphics[width=9truecm]{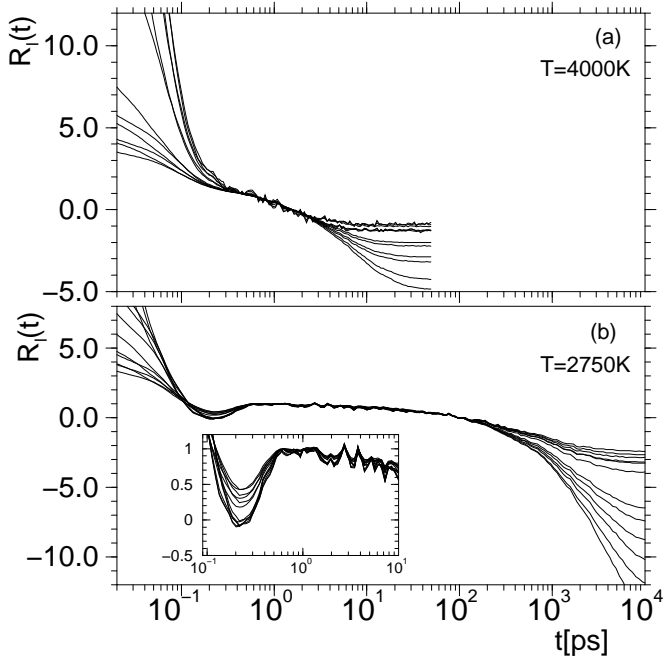}
\caption[]{
Time dependence of the functions $R_l(t)$ as defined in Eq.~(\protect\ref{eq42})
for the case of silica. The upper and lower panel corresponds to $T=4000$~K and
$T=2750$~K, respectively. The correlators used are $F_s(q,t)$ for silicon and
oxygen at $q=1.7$, 2.2, 2.8, 4.43, 5.02 and 5.31~\AA$^{-1}$. The times
$t''$ and $t'$ are 0.4~ps and 1.6~ps for $T=4000$~K and 11~ps and 106~ps for
$T=2750$K, respectively.}
\label{fig_sio2_factorization}
\end{figure}

That MCT is not only able to rationalize some feature of the relaxation
dynamics of the $\alpha-$relaxation but also in the $\beta-$regime is
shown in Fig.~\ref{fig_sio2_factorization}, were we plot the functions
$R_l(t)$ defined in Eq.~(\ref{eq42}). The correlators used to calculate
these curves are the incoherent intermediate scattering function for Si
and O at $q=1.7$, 2.2, 2.8, 4.43, 5.02 and 5.31~\AA$^{-1}$, but other
correlators can be included as well~\cite{kob99b}. As it was the
case of the BLJM system, we see that in the $\beta-$regime the curves
fall on top of each other. In agreement with the prediction of MCT, the
range where a master curve is found expands if the temperature is lowered
(compare the two panels). The shape of the master curve is given by the
$\beta-$correlator $g_-(t)$ from which one can determine the exponent
parameter $\lambda$. Using this value of $\lambda$ (= 0.713), one can
use Eqs.~(\ref{eq30}) and (\ref{eq31}) to calculate the theoretical
value of the critical exponent $\gamma$, which turns out to be 2.35,
in very good agreement with the real value, which is around 2.45.

\begin{figure}
\centering
\includegraphics[width=7truecm]{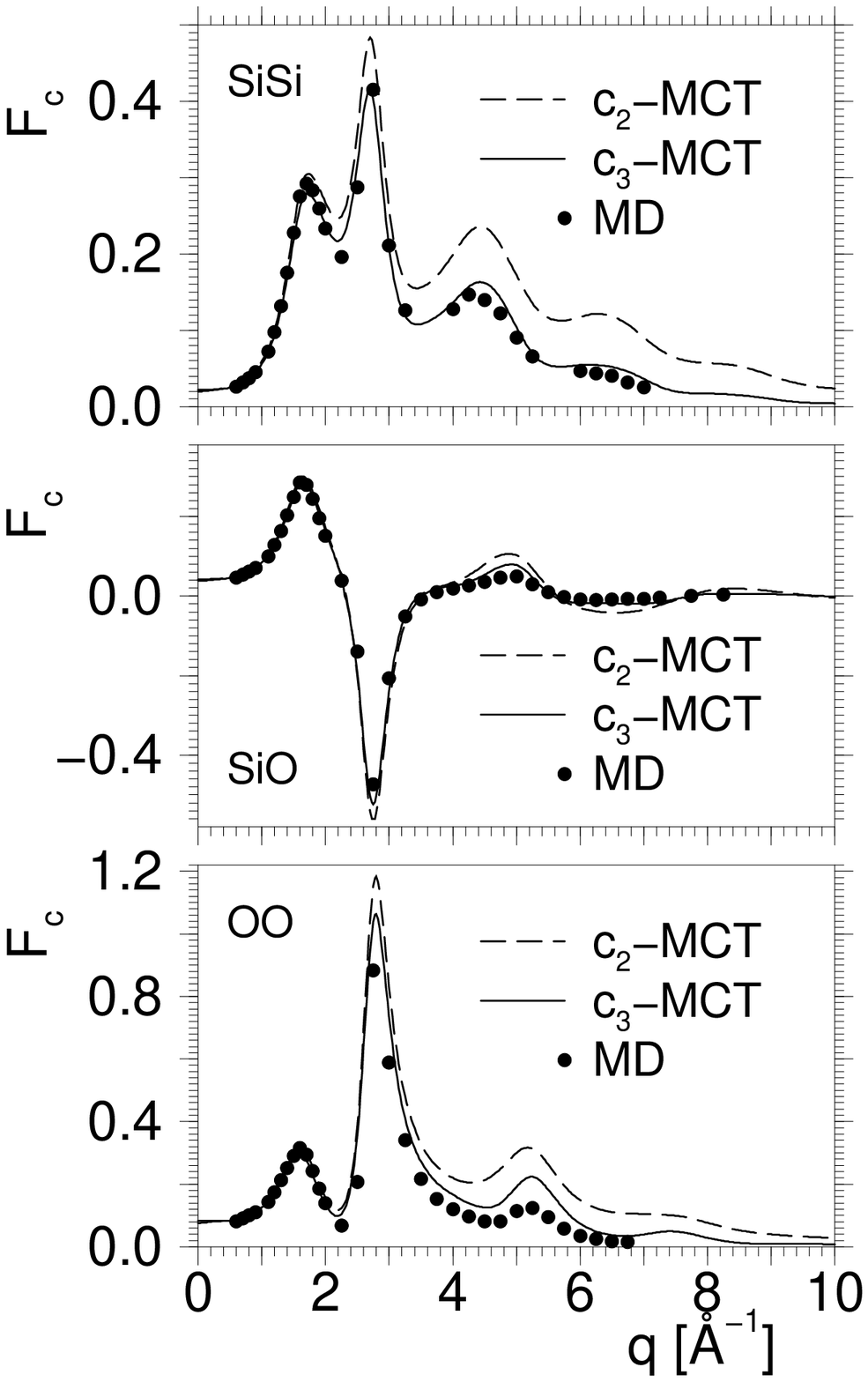}
\caption[]{
Wave-vector dependence of the non-ergodicity parameter for silica.
The dots are the results from the simulations and the dashed and
solid lines are the predictions of MCT neglecting and including the
$c_3-$terms,respectively (see text for details.}
\label{fig_nep_sio2}
\end{figure}

As it was already the case for the BLJM system it is important to
understand also for this network-forming glass whether MCT is able to give
only a qualitative description of the relaxation dynamics, or whether it
is also possible to make a quantitative calculation. It is found that
if one uses the expressions given by Eq.~(\ref{eq20}) for the vertex
$V^{(2)}$ there is no good agreement between, e.g., the wave-vector
dependence of the non-ergodicity as predicted by the theory and the one
found in the simulation~\cite{nauroth99,sciortino01}. This is seen in
Fig.~\ref{fig_nep_sio2} were we compare the result from the simulation
(circles) with the ones of MCT (dashed lines). However, we have already
mentioned in Sec.~\ref{sec3}, that there are also contributions to the
vertices $V^{(2)}$ that are related to the three-particles correlation
function $c_3({\bf q},{\bf k})$. For the case of the BLJM system we
have found, see Fig.~\ref{fig_nep_lj}, that these contributions do not
change significantly the value of the non-ergodicity parameter. For
a network-forming liquid like silica this is, however, no longer the
case in that there is a quite large difference between the theoretical
prediction in which $c_3$ is neglected and the one in which $c_3$ is
taken into account. This can be seen, e.g., in Fig.~\ref{fig_nep_sio2}
were we have included also the curve from the MCT in which the $c_3$ terms
are included (solid line). We find that the curve with and without these
terms differ significantly in that the former is quite a bit lower that
the latter one, in particular at intermediate and large wave-vectors. In
addition we recognize that the inclusion of the inclusion of these terms
lead to a much better agreement between the theory and the data of the
simulation.  For the case of the Si-Si correlation one can even say that
the theory makes a very good prediction for this highly non-trivial
function. Hence we conclude that also for the case of silica, the
prototype of a strong glass-former, the mode-coupling theory is able to
rationalize not only some qualitative features of the relaxation dynamics,
but also to make quantitative predictions that are surprisingly accurate.

\section{Summary and Perspectives}

Space and time limits imposed by the (very wise!) editors, and ignorance
from my side, have had the effect that many highly interesting and
exciting aspects of glassy systems could not be discussed in this
review. The simulations done in the context of the recent theoretical
developments that seem to indicate a much closer connection between
structural glasses, that have no frozen in disorder, and spin glasses, in
which the disorder is quenched, could unfortunately not been mentioned.
Likewise I had to leave out the discussion of the interesting and
promising approach to use the language of ``landscapes'' (potential
and free energy) to describe glassy systems, as well as the fascinating
possibility that also {\it driven} systems (sheared, stirred, etc.) can
be characterized by an effective temperature, thus suggesting a close
connection between these systems and aging glasses. All these topics,
and many others, will certainly be the focus of much research in the
years to come.

Acknowledgments: I warmly thank Jean-Louis Barrat, Mikhail Feigelman,
and Jorge Kurchan for giving me the possibility to give these lectures at
the very interesting and motivating school they organized. In addition I
thank all my friends and colleagues (way too many to be mentioned here)
that have taught me most of the things that I have written about.

\end{document}